\documentclass[floatfix,aip,amsmath,amssymb,preprint]{revtex4-1}

\usepackage{graphicx}%
\usepackage{dcolumn}%
\usepackage{bm}%
\usepackage{mathtools}
\usepackage[utf8]{inputenc}

\usepackage{graphicx}
\usepackage{subcaption}

\usepackage [english]{babel}
\usepackage [autostyle, english = american]{csquotes}

\usepackage{calrsfs}
\DeclareMathAlphabet{\pazocal}{OMS}{zplm}{m}{n}
\SetMathAlphabet\pazocal{bold}{OMS}{zplm}{bx}{n}

\newcommand{\hide}[1]{\ignorespaces}

\usepackage[margin=1in]{geometry}
\usepackage{graphicx}   %
\usepackage{adjustbox}  %
\usepackage{makecell}   %
\usepackage{array}      %
\usepackage{booktabs}

\usepackage[normalem]{ulem}

\usepackage[T1]{fontenc}
\usepackage{mathptmx}
\usepackage{etoolbox}
\usepackage{tikz}

\makeatletter
\def\@email#1#2{%
 \endgroup
 \patchcmd{\titleblock@produce}
  {\frontmatter@RRAPformat}
  {\frontmatter@RRAPformat{\produce@RRAP{*#1\href{mailto:#2}{#2}}}\frontmatter@RRAPformat}
  {}{}
}%
\makeatother

\usepackage{fancyhdr}
\begin{document}

\preprint{AIP/123-QED}

\title[]
{Analyzing Band Gaps in Ensemble Density Functional Theory using Thermodynamic Limits of Finite One-Dimensional Model Systems}
\author{Gregory G. V. Kenning}
\affiliation{Department of Mechanical Engineering and Materials Science, University of Pittsburgh, 4200 Fifth Avenue, Pittsburgh, PA 15260, USA%
}%

\author{Remi J. Leano}

\affiliation{ 
Department of Chemistry and Biochemistry, University of California, Merced, 5200 N. Lake Rd., Merced, CA 95343, USA%
}%

\author{David A. Strubbe}
\affiliation{%
Department of Physics, University of California, Merced, 5200 N. Lake Rd., Merced, CA 95343, USA %
}%
\email{dstrubbe@ucmerced.edu}

\date{\today}%

\begin{abstract}
Ensemble Density Functional Theory (EDFT) is a promising extension to Density Functional Theory (DFT) for calculating excited states. While Kohn-Sham eigenvalue differences underestimate gaps, EDFT has been shown to provide more accurate excitation energies in atoms, molecules and isolated model systems. However, it is unclear whether EDFT is capable of calculating band gaps of periodic systems -- and what an appropriate theoretical formulation would be to describe periodic systems. We explored how EDFT could calculate band gaps by estimating the thermodynamic limit with increasingly wide finite versions of the one-dimensional Kronig-Penney (KP) periodic model. We use Octopus, an \textit{ab initio}, open-source, real-space DFT code, as in our previous work [R. J. Leano \textit{et al.}, Electron. Struct. \textbf{6}, 035003 (2024)] in which we found with ``particle in a box'' models that EDFT can provide a reasonable effective mass correction for the homogeneous electron gas. Now, we use a periodic reference that is gapped. We find that the finite systems' Kohn-Sham gap approaches the same periodic limit for each of three ways of terminating the finite system, though the appropriate states corresponding to the valence band maximum and conduction band minimum have to be carefully identified in each case. Finally, our EDFT results, using a simple ensemblized LDA approximation, have a reasonable nonzero correction to the  bandgap in the periodic limit. The results indicate that EDFT is promising for periodic systems, to motivate further work on developing a suitable formalism.

\end{abstract}
\maketitle

\section{Introduction}
Although Ground State (GS) DFT is incredibly useful in its ability to predict various electronic structure properties, one of its primary issues is underestimation of the optical band gap, \cite{MD87} when calculating it as the difference between the highest occupied and lowest unoccupied Kohn Sham (KS) energy states. Currently, there are multiple techniques within a DFT framework that can be used to improve band gap prediction, including $\Delta$SCF, hybrid functionals, and time-dependent (TD) DFT. \cite{RG84, U12, MMNGR12} Quantum chemistry and GW/Bethe-Salpeter equation approaches can also be used. TDDFT is currently the primary DFT-based method for calculating excitation energies. However, TDDFT has difficulty with double/multiple excitations and periodic systems.\cite{C96} Here, we investigate Ensemble (E) DFT to assess its ability to calculate excitation energies in periodic sysems such as semiconductors. We use the same strategy as our previous work \cite{Remi}, in which we study increasingly large finite systems to approach a periodic limit. Previously, we analyzed increasingly wide finite 1D ``particle in a box'' systems, which approach the homogeneous electron gas in the periodic limit. We found that the gap correction (and gap) approached zero, and inferred a reasonable effective mass from the corrections for finite gaps. These results indicate that EDFT is promising for metallic systems. In this work, using instead a model periodic system with a nonzero gap, we can study how the gap itself is corrected in the periodic limit, to relate to semiconductors and insulators. 

The Kronig-Penney model \cite{KP} is a periodic one-dimensional model with periodic finite rectangular barriers evenly spaced between rectangular wells, as shown in Figure \ref{fig:schematic_potentials}, and is one of the simplest models that has a band gap. (A version with Dirac $\delta$ barriers is also sometimes studied.) Successful treatment of the band gap of a Kronig-Penney model with EDFT would indicate promise for the application of EDFT on real semiconductors. Previous literature on finite KP models have discussed variable properties such as potential heights and barrier widths,\cite{VarP} interactions of the finite KP model with lasers, \cite{laser} topological states, \cite{top} surface states induced by the presence of a different potential at the surface, \cite{Coating} and tunneling between finite KP barriers. \cite{Binding} The approach of these systems to the infinite limit, and the electronic structure of systems with interacting electrons, are not found in the literature however. 
\section{Theory}
\subsection{\label{sec:KP_model}Kronig-Penney Model}

\begin{figure}[htbp]
    \centering
    \caption{Potentials for the Kronig--Penney model: periodic case and three finite variants with different centerings relative to the domain edges.}
    \label{fig:schematic_potentials}

    \begin{subfigure}{0.24\textwidth}
        \centering
        \includegraphics[width=\linewidth]{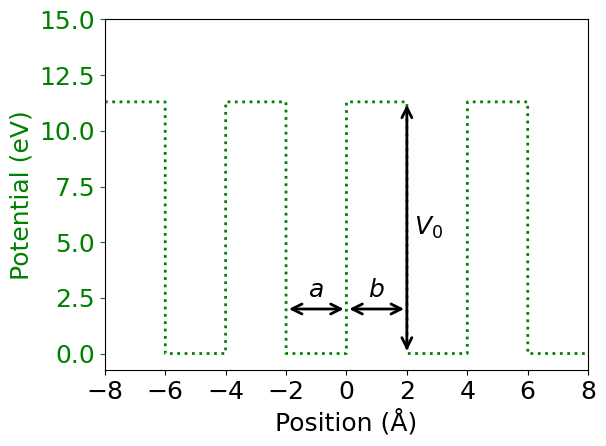}
        \caption{Periodic}
    \end{subfigure}
    \hfill
    \begin{subfigure}{0.24\textwidth}
        \centering
        \includegraphics[width=\linewidth]{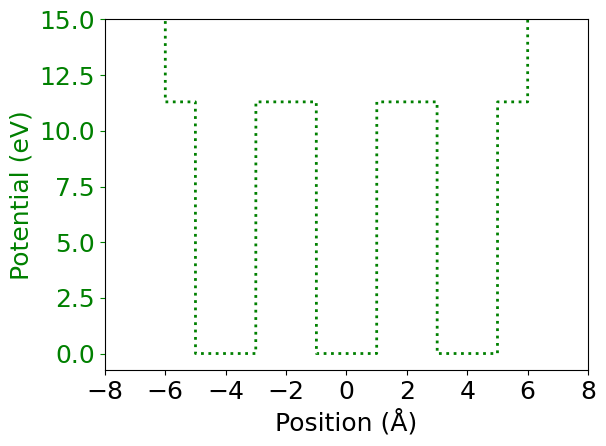}
        \caption{Well-centered}
    \end{subfigure}
    \hfill
    \begin{subfigure}{0.24\textwidth}
        \centering
        \includegraphics[width=\linewidth]{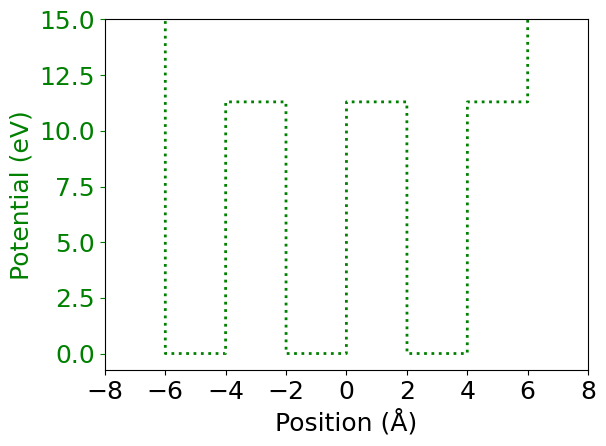}
        \caption{Edge-centered}
    \end{subfigure}
    \hfill
    \begin{subfigure}{0.24\textwidth}
        \centering
        \includegraphics[width=\linewidth]{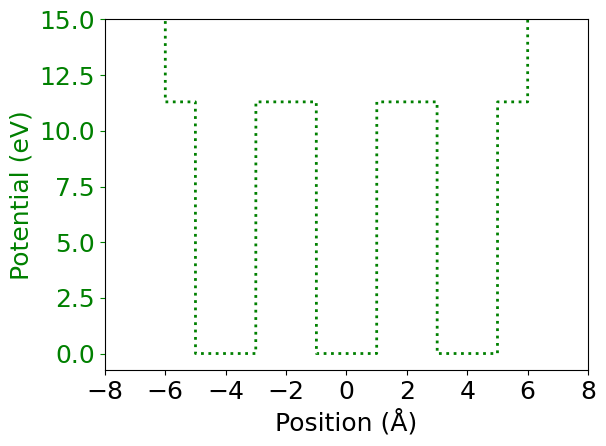}
        \caption{Barrier-centered}
    \end{subfigure}

\end{figure}

\begin{figure}[h]
\centering
\includegraphics[width=.8\columnwidth]{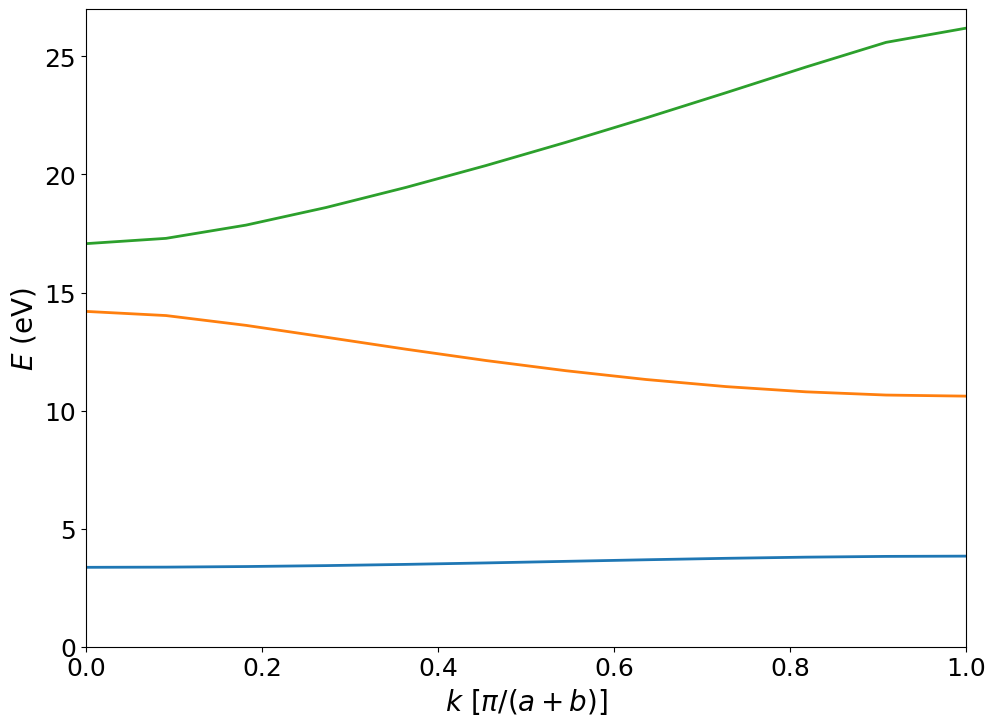} 
\caption{Bandstructure of a Kronig-Penney model with potential barrier widths and wells $a = b = 2.0$ \AA \ and potential height $V_0 = 11.3$ eV. This model has bandwidths of 0.474 eV and 3.589 for the first two bands.}
\label{fig:band}
\end{figure}

We define a KP model with a length of 2.0 Å for both the width of the wells and the barrier, and  a potential height of 11.3 eV (Fig. \ref{fig:schematic_potentials}). This produces a band gap near 6.78 eV, which is in the range for a wide-bandgap semiconductor or insulator. The bandstructure of this model is shown in Fig. \ref{fig:band}. 

We make a finite version of the KP model by considering a certain number of unit cells, treated in a finite domain with zero boundary conditions for the wavefunction, which may also be thought of as the potential going to infinity outside the domain.
We consider three flavors of our finite KP system: (1) centered on the center of one of the wells (a ``well-centered system''), so that the system is terminated in the middle of the barrier; (2) centered on the discontinuity in the potential between well and barrier (an ``edge-centered system''), so that the system is terminated by a well on one side and a barrier on the other; and (3) centered on the center of one of the barriers (a ``barrier-centered'' system), so that the system is terminated in the middle of a well. These options of ``centerings'' are depicted in Fig. \ref{fig:schematic_potentials}.

\subsection{Ensemble Density Functional Theory}
Gross-Oliveira-Kohn (GOK) generalized Theophilou and Gidopoulos's variational principle for equi-ensembles built from KS states to ensembles of monotonically decreasing, non-equal weights.  \cite{T87,GOK88I} We consider an ensemble of $M_{I}$ $n$-electron states, which may be degenerate, numbered from $m=0$ to $M_{I}-1$. We label the multiplicity of the $I$th multiplet $g_I$.\cite{GOK88I, MSFL20} $I$ is the highest-energy set of degenerate states in the ensemble, which we call the $(M_{I}-1)$th state. All of each degenerate subspace must be included in the GOK ensemble in order for the ensemble to be well-defined. Considering the non-interacting single-particle wavefunctions $\varphi^w_{j}$, we construct the non-interacting ensemble KS equation:
\begin{equation}
\left\{-\frac{1}{2}\nabla^2 + v^w_{\text{KS}}[\rho^w](r)\right\}\varphi^w_{j}(r) = \epsilon^w_{j}\varphi^w_{j}(r).
\label{eqn:ensKS}
\end{equation}
From the ensemble KS equation, equation (\ref{eqn:ensKS}), individual energies $\epsilon_j$ are obtained from $\{\varphi_j(r)\}_{1 \leq j \leq \infty}$. The wavefunctions $\varphi^w_{j}$ are defined such that they reproduce the ensemble density:
\begin{equation}
\rho^w(r) = \sum^{M_{I}-1}_{m=0} \mathtt{w}_m \bigg(\sum^\infty_{j=1} f^{m}_{j}|\varphi_j(r)|^2 \bigg),
\label{eqn:ensdens}
\end{equation}
where the occupation of  $\varphi_j(r)$ in the $m$th KS wave function $\Psi^w_m[\rho^w]$ is denoted by $f^m_{j}$.\cite{MSFL20} The KS many-body wavefunctions $\{\Psi_m^{w}[\rho]\}_{0 \leq m \leq M_{I}-1}$ are built from $\{\varphi_j(r)\}_{1 \leq j \leq \infty}$ and are assumed to be Slater determinants, or linear combinations of Slater determinants. The Kohn-Sham potential is given by
\begin{equation}
    v_{\text{KS}}[\rho](x,\sigma) = v_{\text{ext}}(x) + e^2 \int \frac{ \sum_{\sigma'=\alpha}^\beta \rho(x',\sigma')}{|x-x'|}dx' + \frac{\delta E_{\text{xc}}[\rho]}{\delta \rho(x,\sigma)},
    \label{eqn:vs}
\end{equation} 
which can be written in an ensemble-generalized form as
\begin{equation}
v^w_{\text{KS}}[\rho^w](r) = v_{\text{ext}}(r) + \frac{\delta E^w_{\rm Hxc}[\rho^w]}{\delta \rho^w(r)}.
\end{equation}
The ensemble functional for Hartree, exchange, and correlation (HXC), $E^w_{\rm Hxc}$, is written as a sum of each:
\begin{equation}
    \frac{\delta E^w_{\rm Hxc}[\rho^w]}{\delta \rho^w(r)} = \int \frac{\rho^w(r')}{|r-r'|}dr' + \frac{\delta E_{\rm x}^w[\rho^w]}{\delta \rho^w(r)} + \frac{\delta E_{\rm c}^w[\rho^w]}{\delta \rho^w(r)}.
\end{equation}
The energy of the $m$th many-electron KS state is
\begin{equation}
E_m = \sum^\infty_{j=1} f^{m}_{j} \epsilon_j,
\label{eqn:EmthKS}
\end{equation}
with $m$ an integer $0 \leq m \leq M_{I-1}$. Each $\epsilon_j$ is the non-interacting (Kohn-Sham) energy for a state in the system. From the set $\{w\} \equiv (\mathtt{w}_{m=0}, ..., \mathtt{w}_{m=M_{I}-1})$ of monotonically non-increasing $(\mathtt{w}_{m=0} \geq ... \geq \mathtt{w}_{m=M_{I}-1})$ weights obeying
\begin{equation}
    \sum_{m=0}^{M_{I}-1} \mathtt{w}_m = 1,
\end{equation}
we assign each state a weight $\mathtt{w}_m$. 
This requirement of a discrete set of states is the reason this theory cannot be used directly for a periodic system, which is a continuum of states. The weights of the ensemble are defined as as typically done for a GOK-I ensemble\cite{GOK88I}
\begin{equation}
\mathtt{w}_m =
    \begin{cases}
            \frac{1-\mathtt{w}g_I}{M_{I} - g_I} & m < M_{I} - g_I,   \\
            \mathtt{w} &   m \geq M_{I} - g_I, 
    \end{cases}
    \label{eqn:GOKweights}
\end{equation}
where $\mathtt{w} \in [0, 1/M_{I}]$. The weights for all states is equal, except for those in the $I$th multiplet. By this setup, only the weight of the states in the $I$th multiplet, $\texttt{w}$, is needed to define the weights for the entire ensemble. 
The excitation energy $\Omega_I$ of multiplet $I$ from the GS is obtained by a differential of the total ensemble energy with respect to $\mathtt{w}$:\cite{DF19}

\begin{equation}
    \Omega_{I} = E_I - E_0 + \left. \frac{\partial E_{\text{Hxc}}^w [\rho]}{\partial \texttt{w}} \right|_{\rho=\rho^w} .
    \label{eqn:omega_I}
\end{equation}
The \enquote{ensemble correction} to the non-interacting difference of energies between the $I$th KS state and the GS is provided by the last term on the right.

\section{Numerical Calculations}
We start with a spin-polarized independent particle calculation for 1D finite systems with 2 electrons per unit cell. The calculations are performed in the Octopus real-space DFT code, which can straightforwardly handle 1D systems and model systems.\cite{Octopus_PCCP,Octopus_2020}  %
We used ground state self-consistent field (SCF) calculations as the input for further "one-shot" calculations which begin from the ground-state density. In order to analyze higher states, we calculated 10 extra states. We define convergence as when all states have individual errors less than 1e-7 for two consecutive iterations in the SCF loop. We do not use any preconditioner to solve the Kohn-Sham equations. In order to achieve swift convergence, we set no limit on the maximum number of SCF iterations for non-interacting calculations but limit the maximum number of eigen calculation iterations to 50. The radius allows us to define the size of our finite system, so while creating a function like this and defining the variable n, the number of electrons in our system, allows us to treat each new run as a function of the number of electrons that we are analyzing. The spacing is the largest reasonable spacing between data points that converges to the correct value. These runs typically converged anywhere between 300 and 900 iterations, allowing for runs to finish within reasonable amounts of time while converging to results within 1 meV of eachother. 

For a given value of the number of electrons, we run an independent particles, or Kohn Sham, calculation. Next, we  do a one-shot calculation which uses the wavefunctions from the independent particle calculation, but occupations of the KS states for each state in the ensemble are built based on Slater determinants from Ref. \cite{Remi} in order to obtain $E_{\rm H}$, $E_{\rm x}$, and $E_{\rm c}$ for a density built from the given occupations. %

\section{Results}
\subsection{Non-Interacting System}
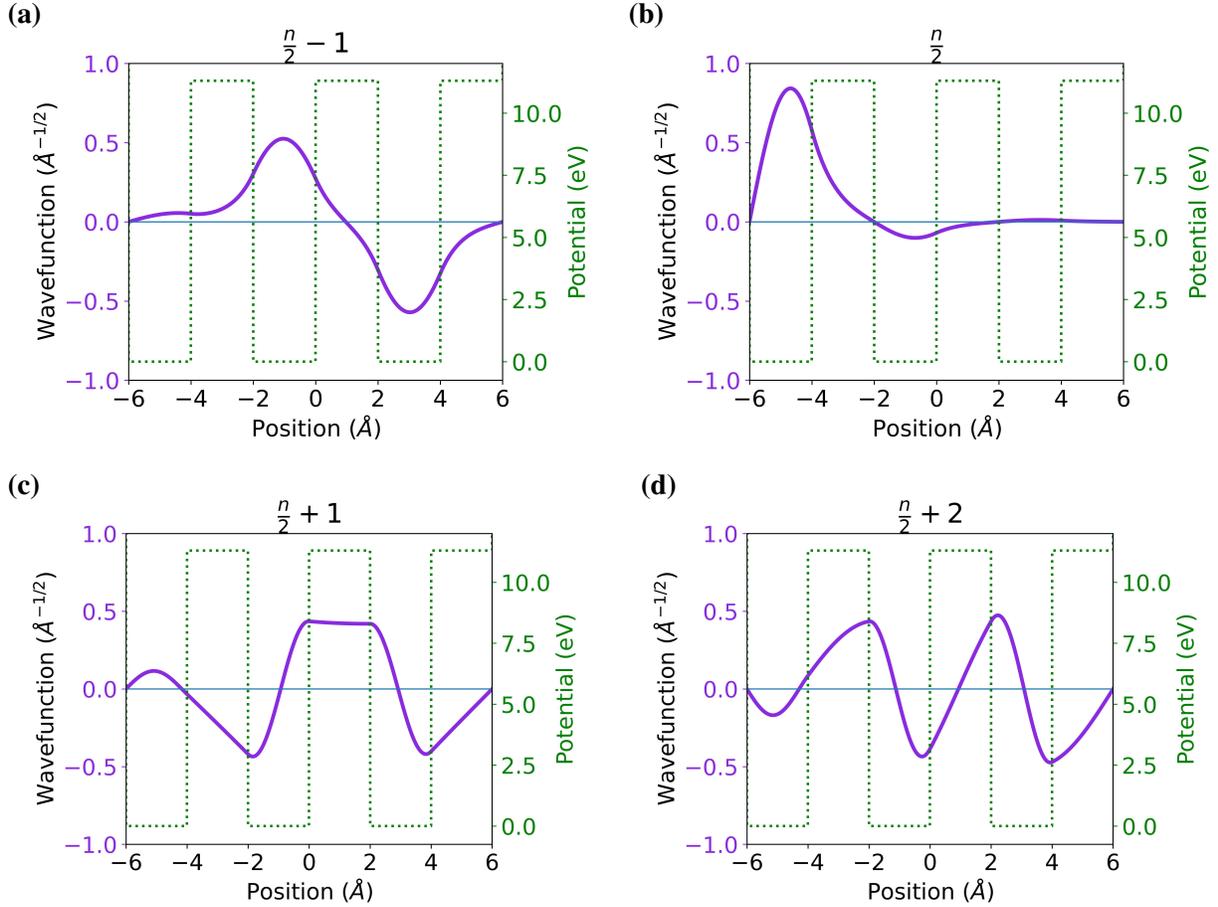
\begin{figure*}[ht]	
\centering{
\begin{tikzpicture}
\node [anchor=north west] (imgA) at (-0.10\linewidth,.58\linewidth){\includegraphics[width=0.49\linewidth]{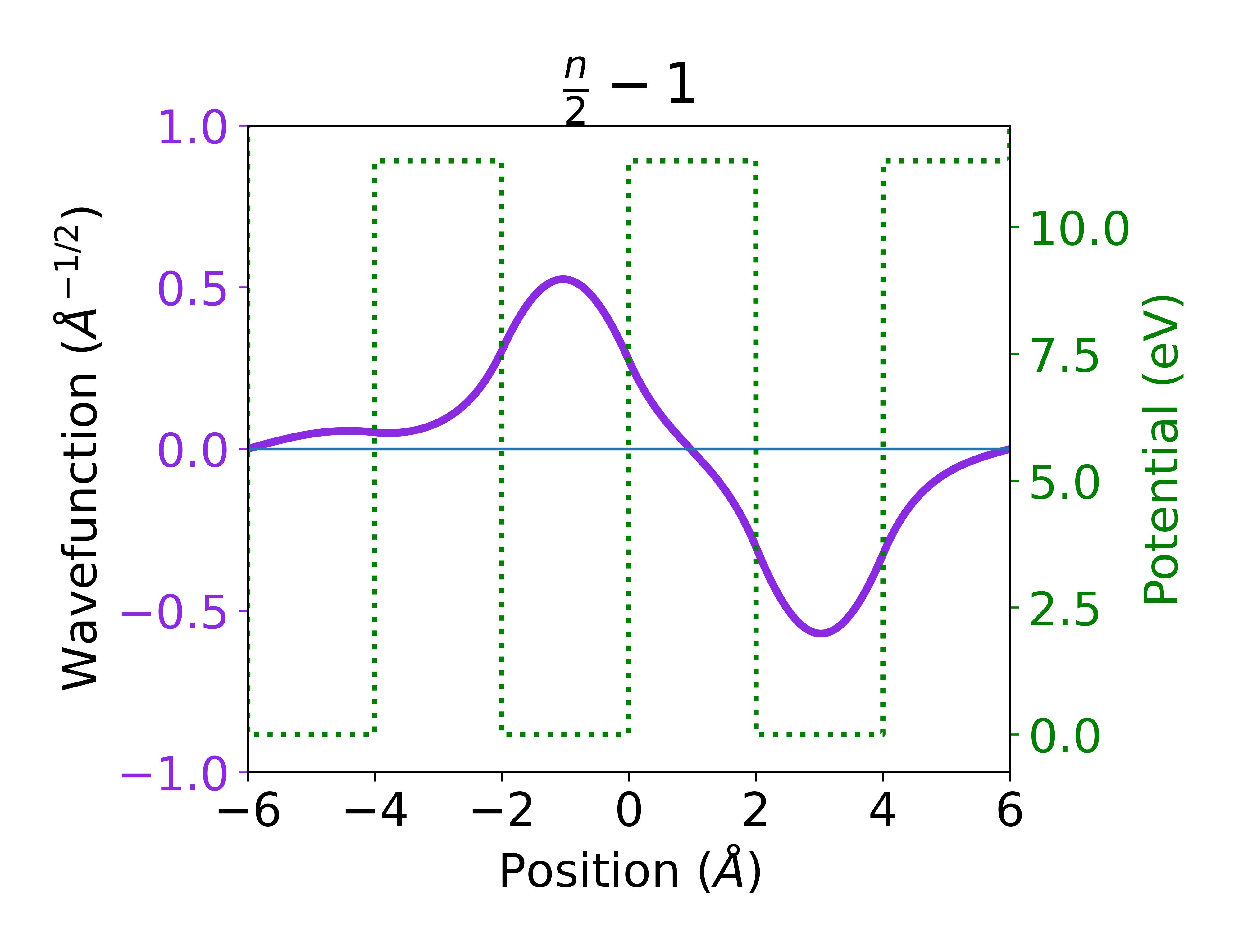}};
             
\node [anchor=north west] (imgB) at (0.40\linewidth,.58\linewidth){\includegraphics[width=0.49\linewidth]{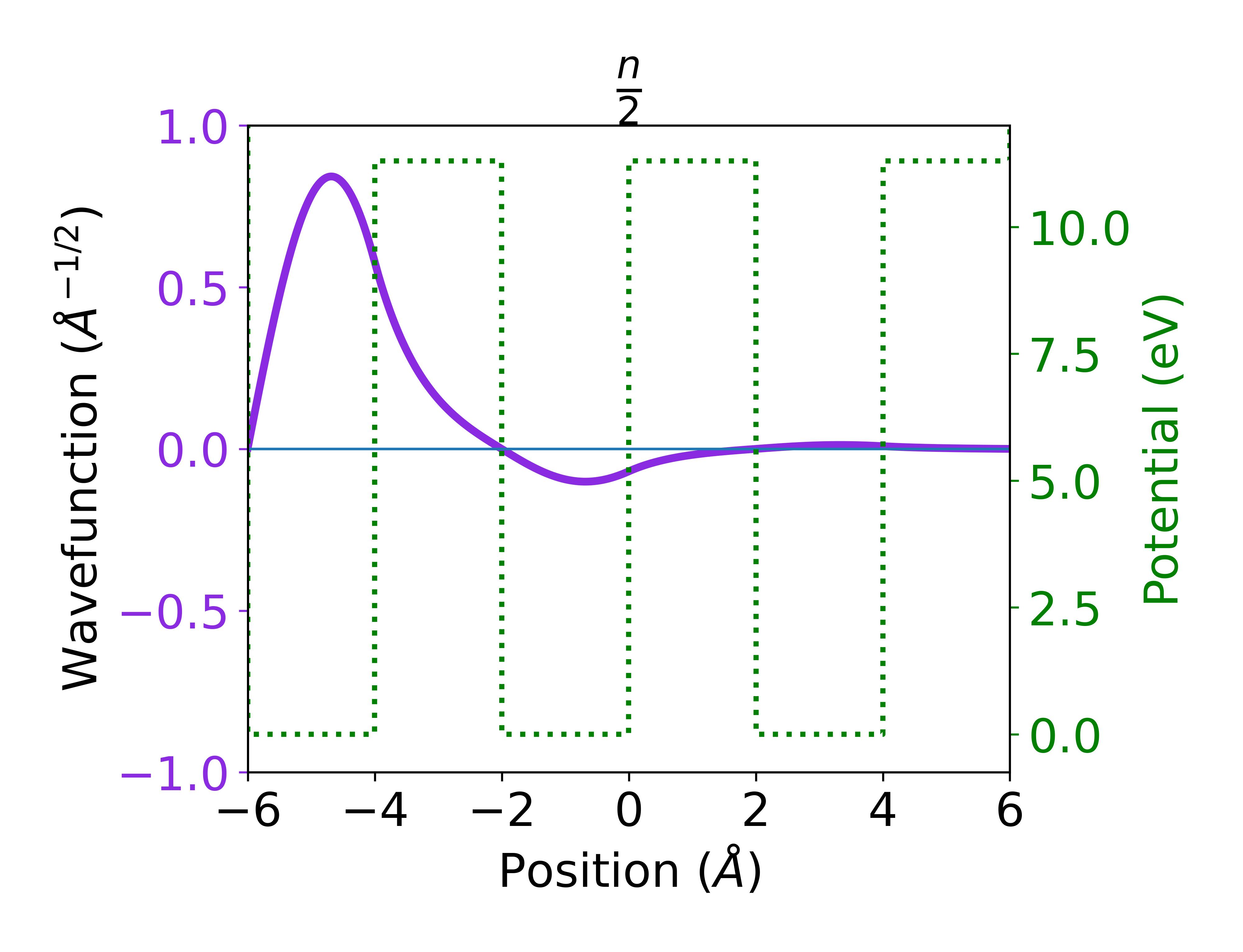}};    
          
\node [anchor=north west] (imgC) at (-0.10\linewidth,.2\linewidth){\includegraphics[width=0.48\linewidth]{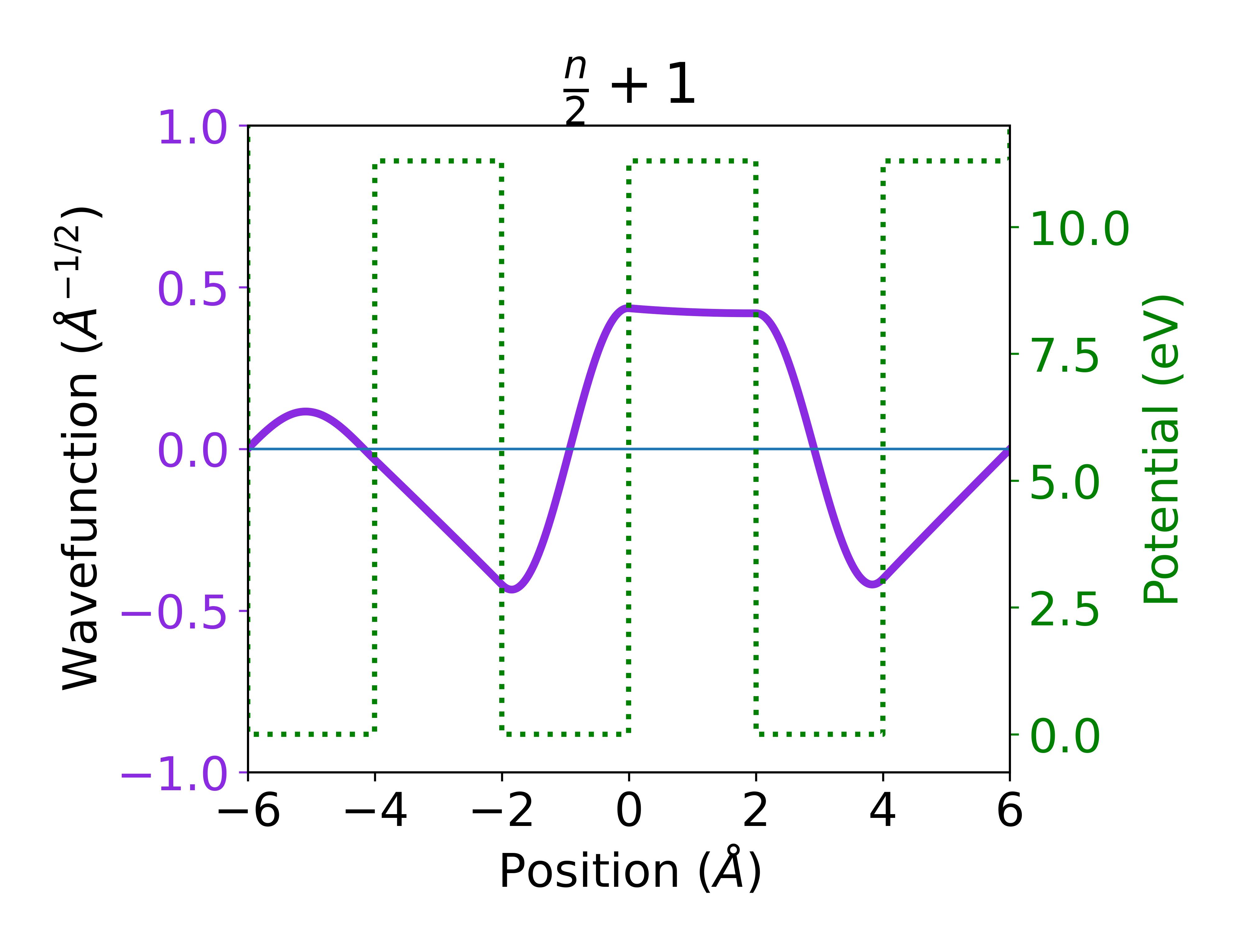}};

\node [anchor=north west] (imgD) at (0.40\linewidth,.2\linewidth){\includegraphics[width=0.48\linewidth]{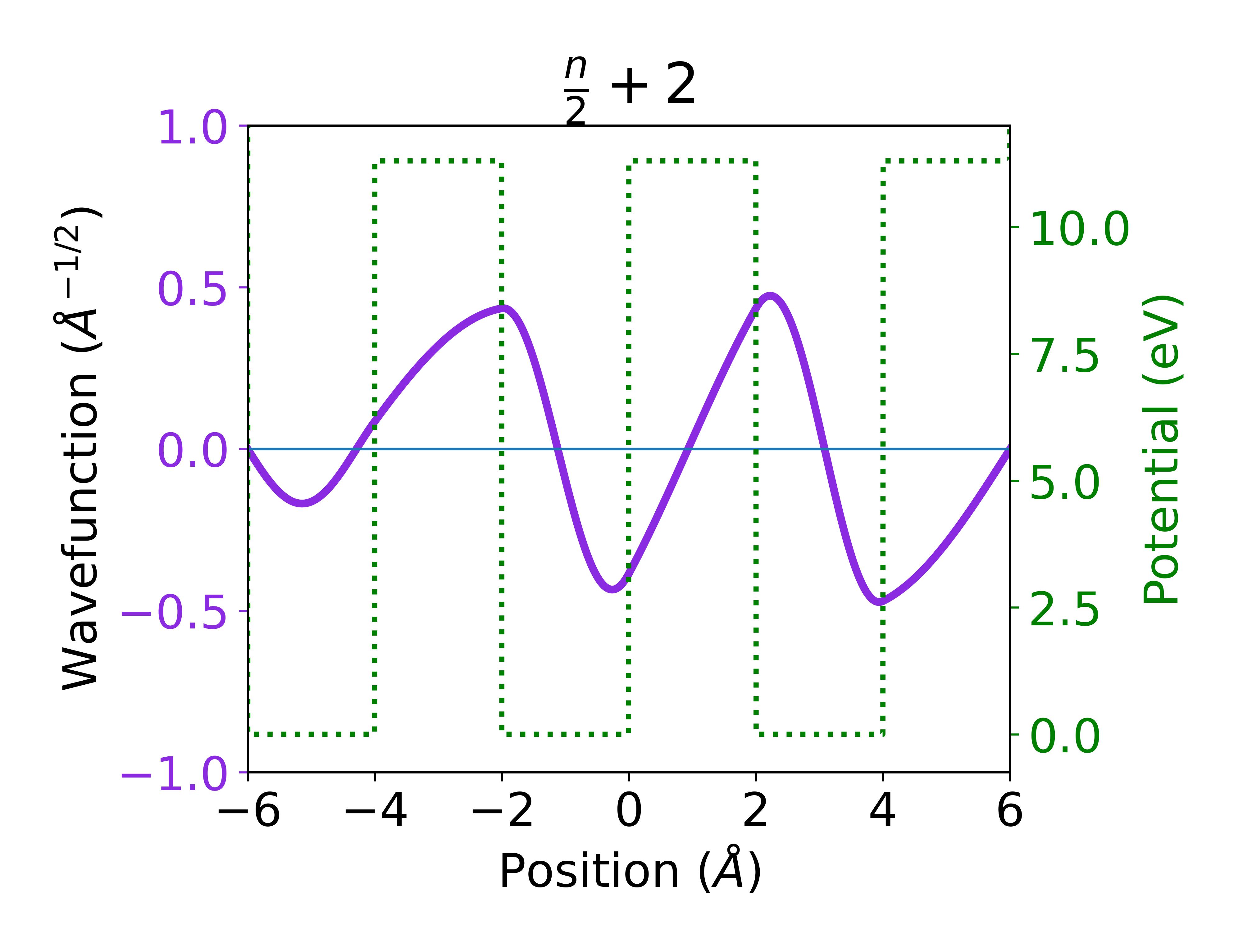}};
       
\draw [anchor=north west] (-0.10\linewidth, .58\linewidth) node {\textbf{(a)} };
\draw [anchor=north west] (0.40\linewidth, .58\linewidth) node {\textbf{(b)} };
\draw [anchor=north west] (-0.10\linewidth, .2\linewidth) node {\textbf{(c)} };
\draw [anchor=north west] (0.41\linewidth, .2\linewidth) node {\textbf{(d)} };

\end{tikzpicture}}
\caption{Wave functions for Edge-Centered System (6e)}
\label{WF-edge}
 \end{figure*}

\begin{figure}[h]
\centering
\includegraphics[width=.8\columnwidth]{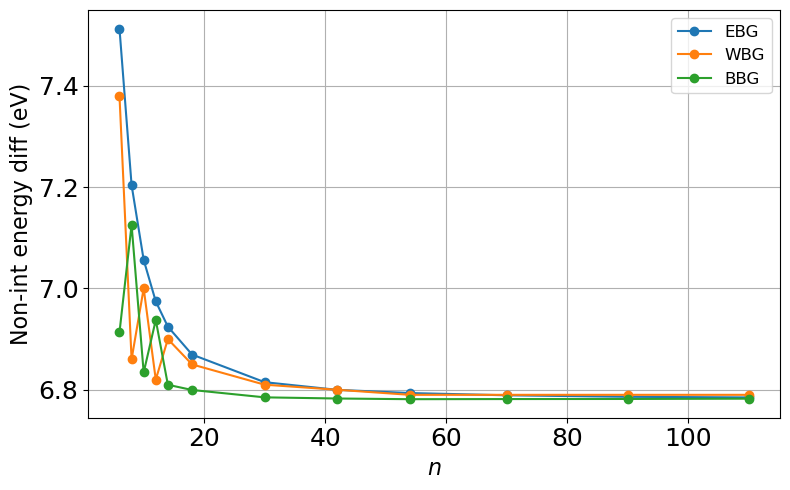} 
\caption{Bandgaps approaching periodicity for non-interacting all centerings. With EBG representing Edge centered bandgap, BBG representing Barrier centered bandgap, WBG representing Well centered bandgap.}
\label{ABGVN}

\end{figure}

\begin{figure}[h]
\centering
\includegraphics[width=.8\columnwidth]{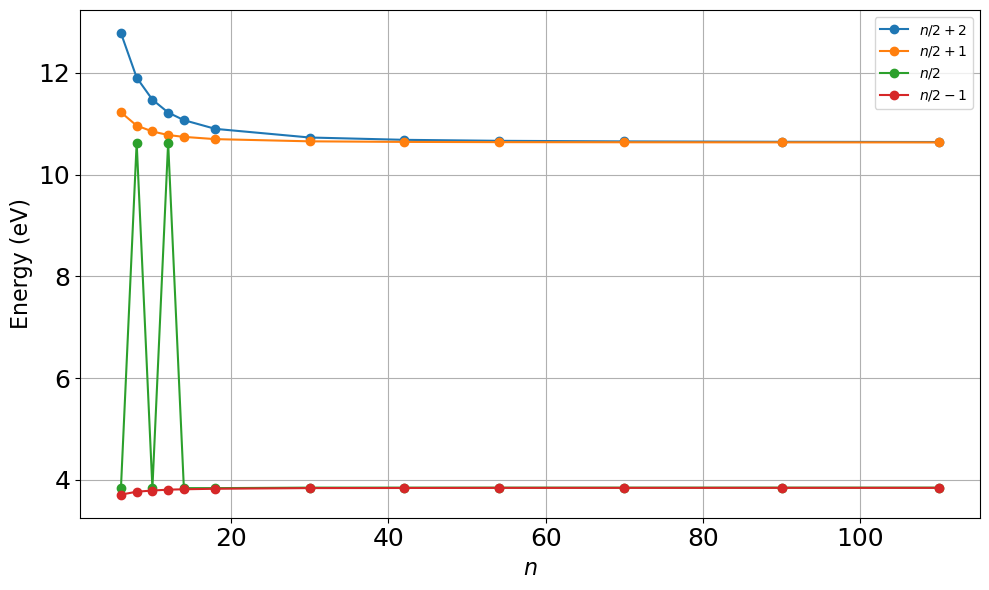} 
\caption{State energies of the approach to periodicity for non-interacting system in well centered system. Degeneracy occurs when switching between the two electron count patterns. }
\label{WellStates}
\end{figure}

For our initial results, we first analyzed the non-interacting particle data. We found that as $N_\text{e}$ increases for the non-interacting system, the bandgap slowly approaches that of the periodic system, as shown in Fig. \ref{ABGVN}. We also found that the the bandgap calculation is dependent on the number of nodes for a given centering, rather than electron count. As shown in figure \ref{WellStates}, the $n/2$ state switches from being the HOMO to the LUMO state when switching from a base $4m +2$ electron count to the $4m$ electron count. In doing this we found that the three different possible centerings for the system, described in Section \ref{sec:KP_model}, that all approach in slightly different bandgaps in the limit. Depending on this choice of centering, different states need to be used to find the bandgap for the system. For the edge centered system, considering an electronic state indexed by $n$, one needs to subtract the $n/2+1$ state from the $n/2-1$ state, as can be seen in Figure \ref{WF-edge}. This results in a periodic bandgap of 6.784 eV. For the edge centering, these two states contain 3 and 1 nodes respectively, which is where the bandgap occurs for this centering no matter the choice of interval in number of electrons in the increase towards the limit. When looking at the wave functions for the states in this system, one finds that they normalize a consistent amount of times for certain states in each centering to find the same bandgap. One will also find however the inclusion of an edge state. This edge state $n/2$ is the cause of the required skipping of a state to find the true bandgap in the system, and is exclusive to the edge centered system. 

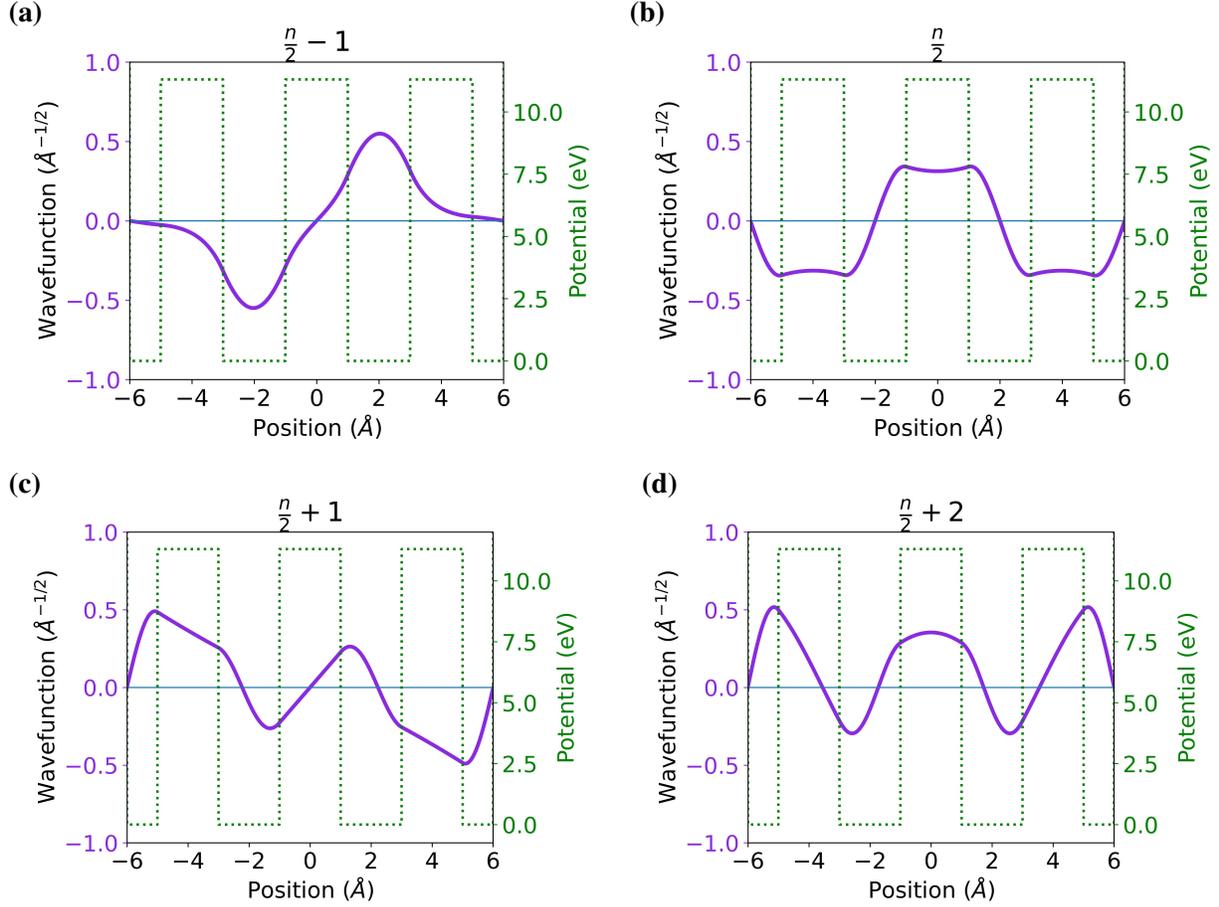
\begin{figure*}[ht]	
\centering{
\begin{tikzpicture}
\node [anchor=north west] (imgA) at (-0.10\linewidth,.58\linewidth){\includegraphics[width=0.49\linewidth]{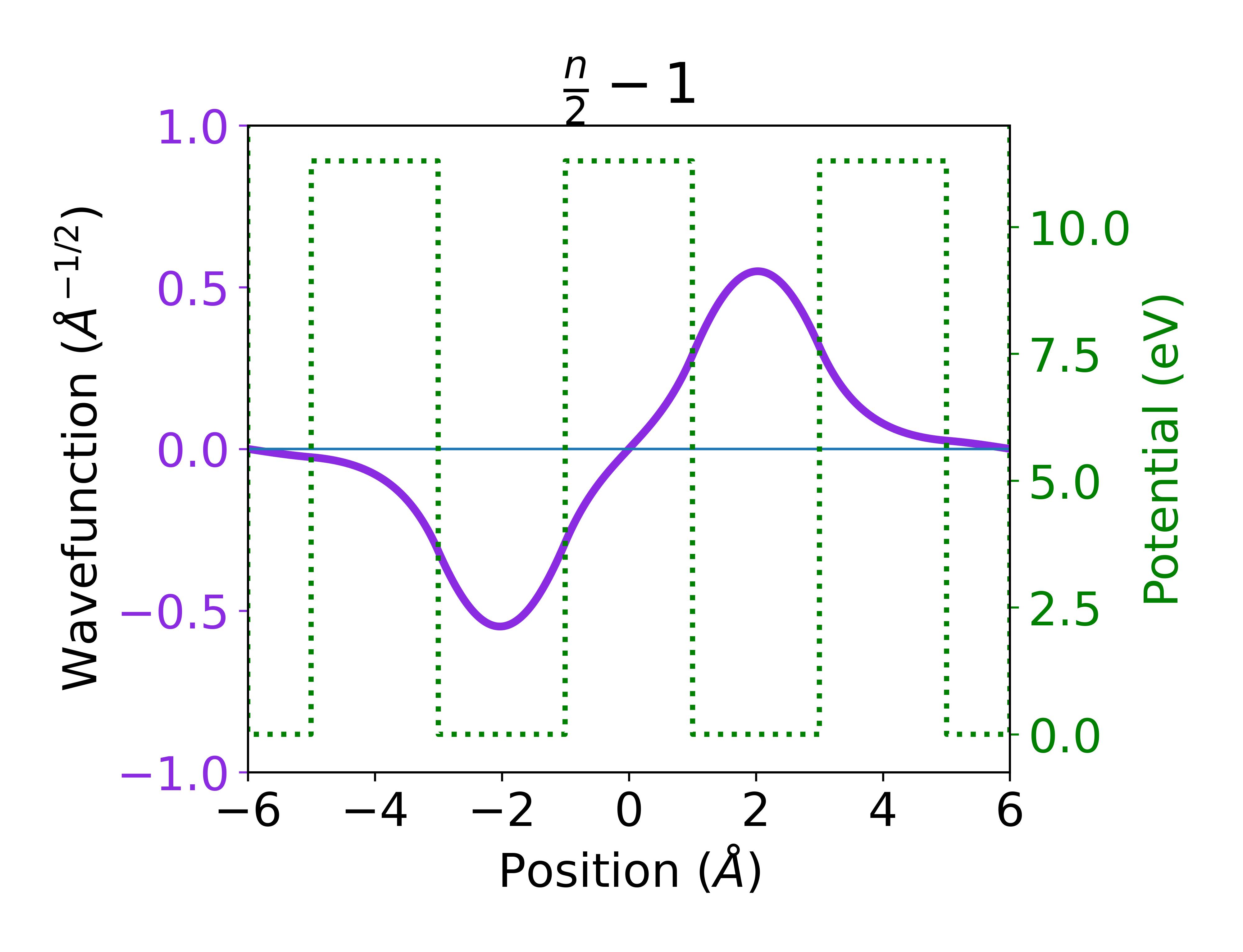}};
\node [anchor=north west] (imgB) at (0.40\linewidth,.58\linewidth){\includegraphics[width=0.49\linewidth]{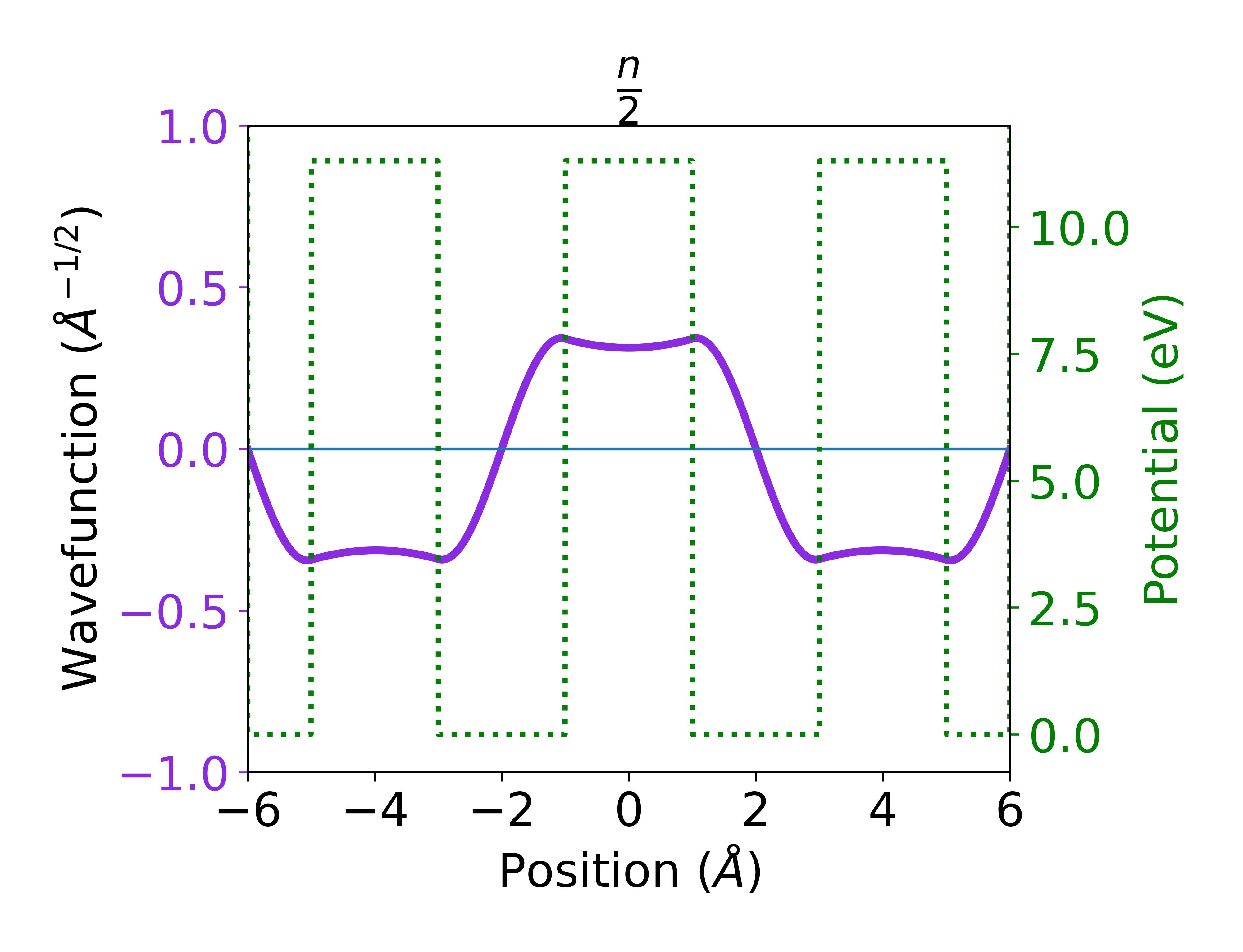}};    
\node [anchor=north west] (imgC) at (-0.10\linewidth,.2\linewidth){\includegraphics[width=0.48\linewidth]{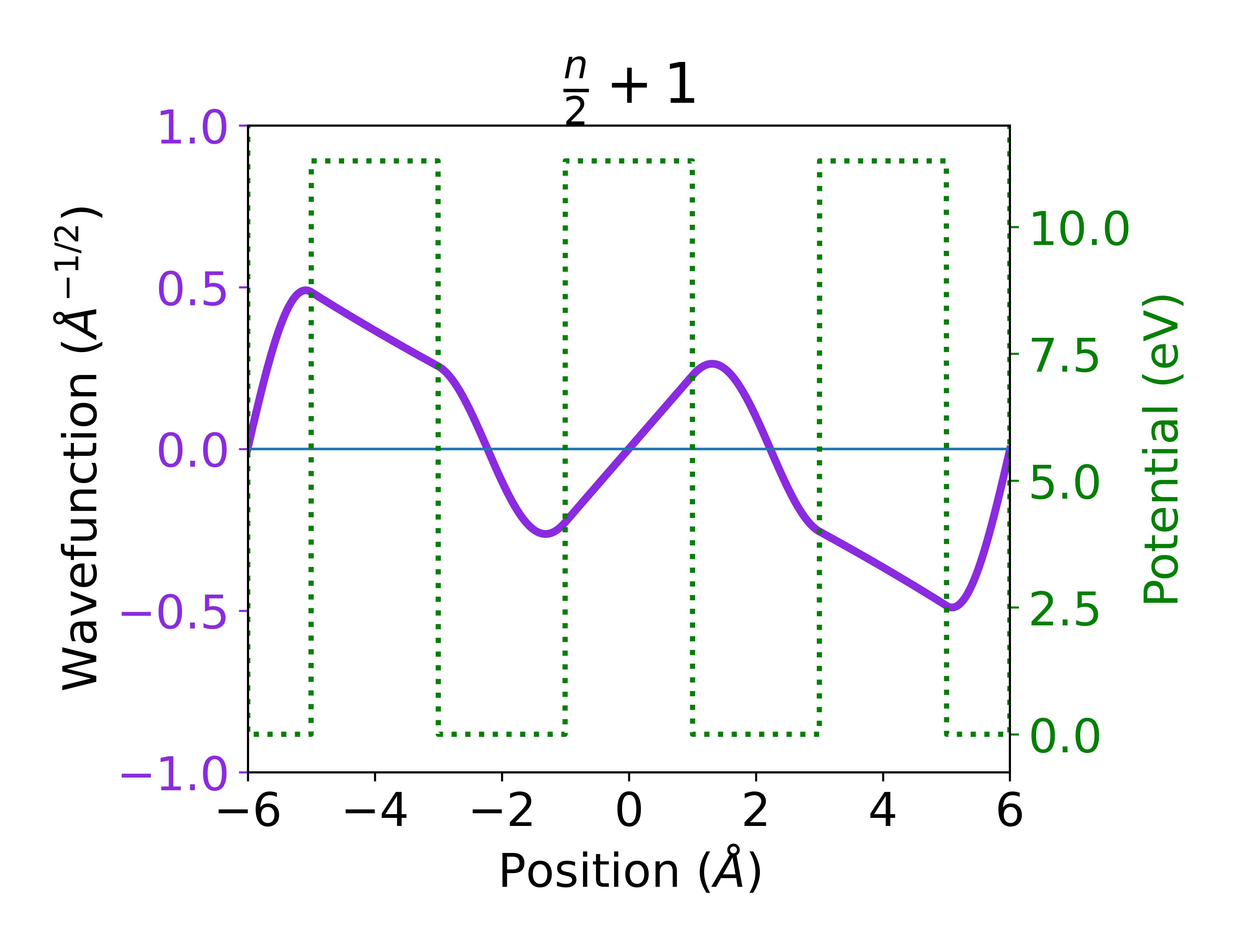}};
\node [anchor=north west] (imgD) at (0.40\linewidth,.2\linewidth){\includegraphics[width=0.48\linewidth]{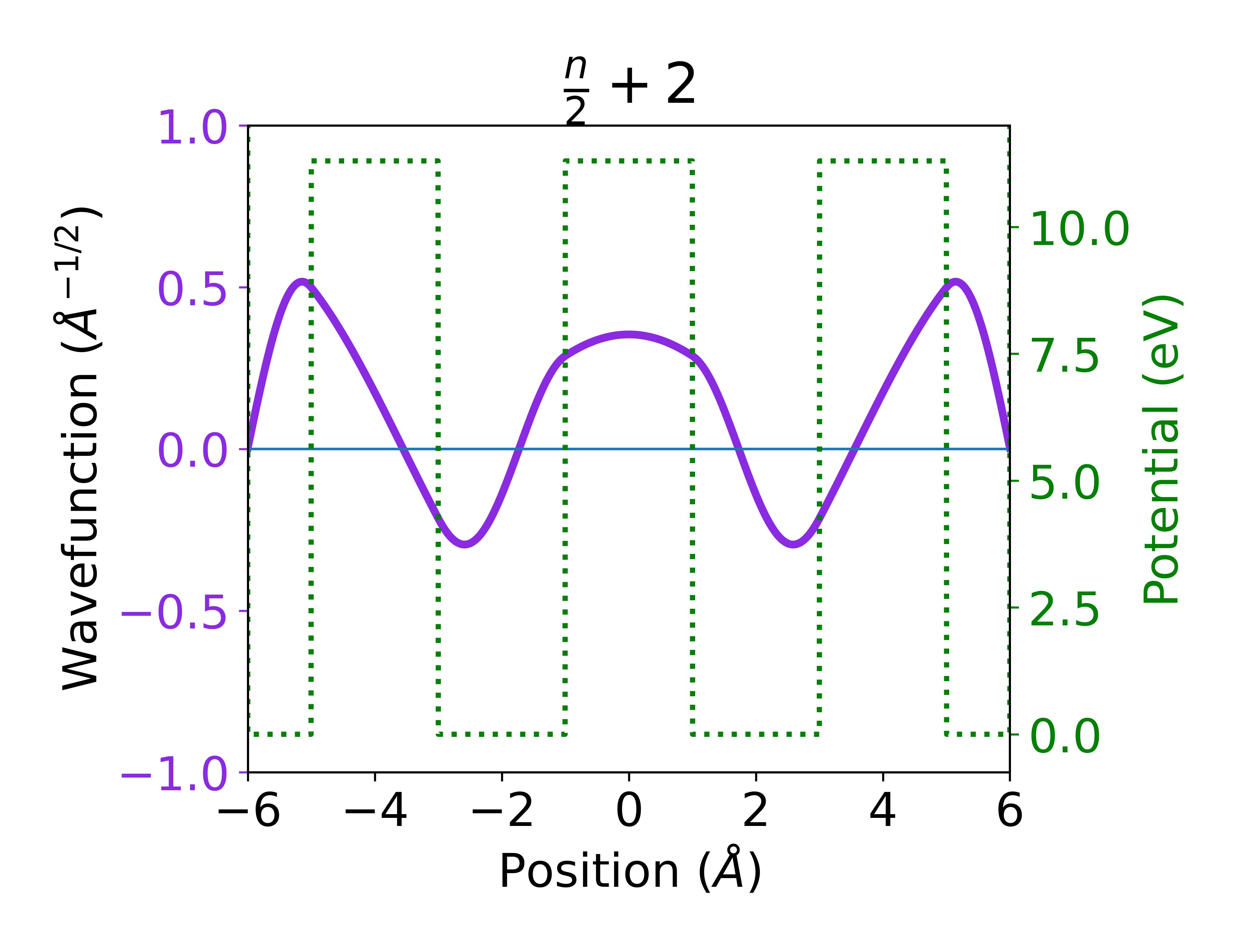}};
\draw [anchor=north west] (-0.10\linewidth, .58\linewidth) node {\textbf{(a)} };
\draw [anchor=north west] (0.40\linewidth, .58\linewidth) node {\textbf{(b)} };
\draw [anchor=north west] (-0.10\linewidth, .2\linewidth) node {\textbf{(c)} };
\draw [anchor=north west] (0.41\linewidth, .2\linewidth) node {\textbf{(d)} };
\end{tikzpicture}}
\caption{Wave functions for Barrier-Centered System (6e)}
\label{WF-Barrier}
 \end{figure*}

For the barrier-centered system the bandgap occurs when using the $n/2$ state and the $n/2-1$ state, which contain 2 and 1 nodes respectively. These states can be seen in figure \ref{WF-Barrier}. When subtracting these states from each other, a bandgap of 6.773 eV is obtained in the approach to the limit.

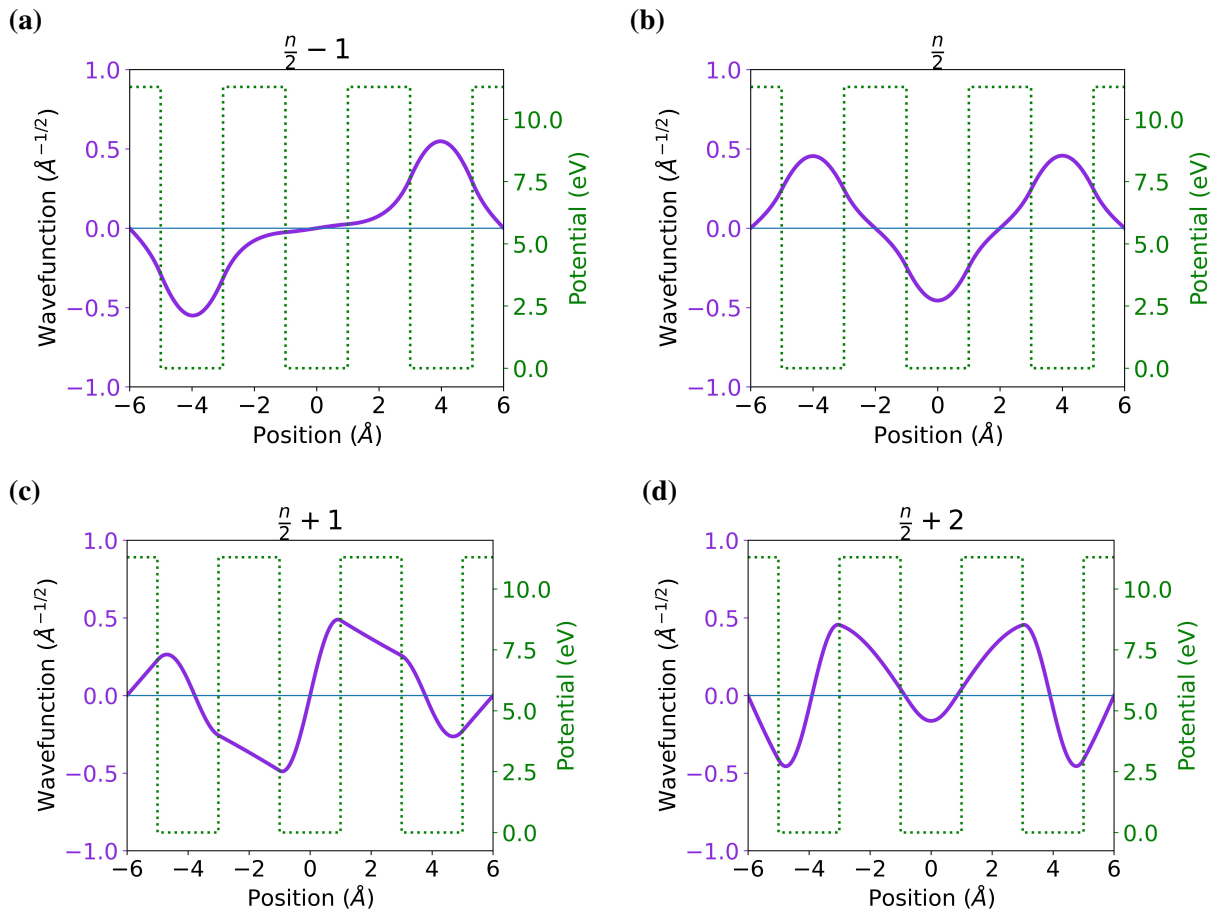
\begin{figure*}[ht]	
\centering{
\begin{tikzpicture}
\node [anchor=north west] (imgA) at (-0.10\linewidth,.58\linewidth){\includegraphics[width=0.49\linewidth]{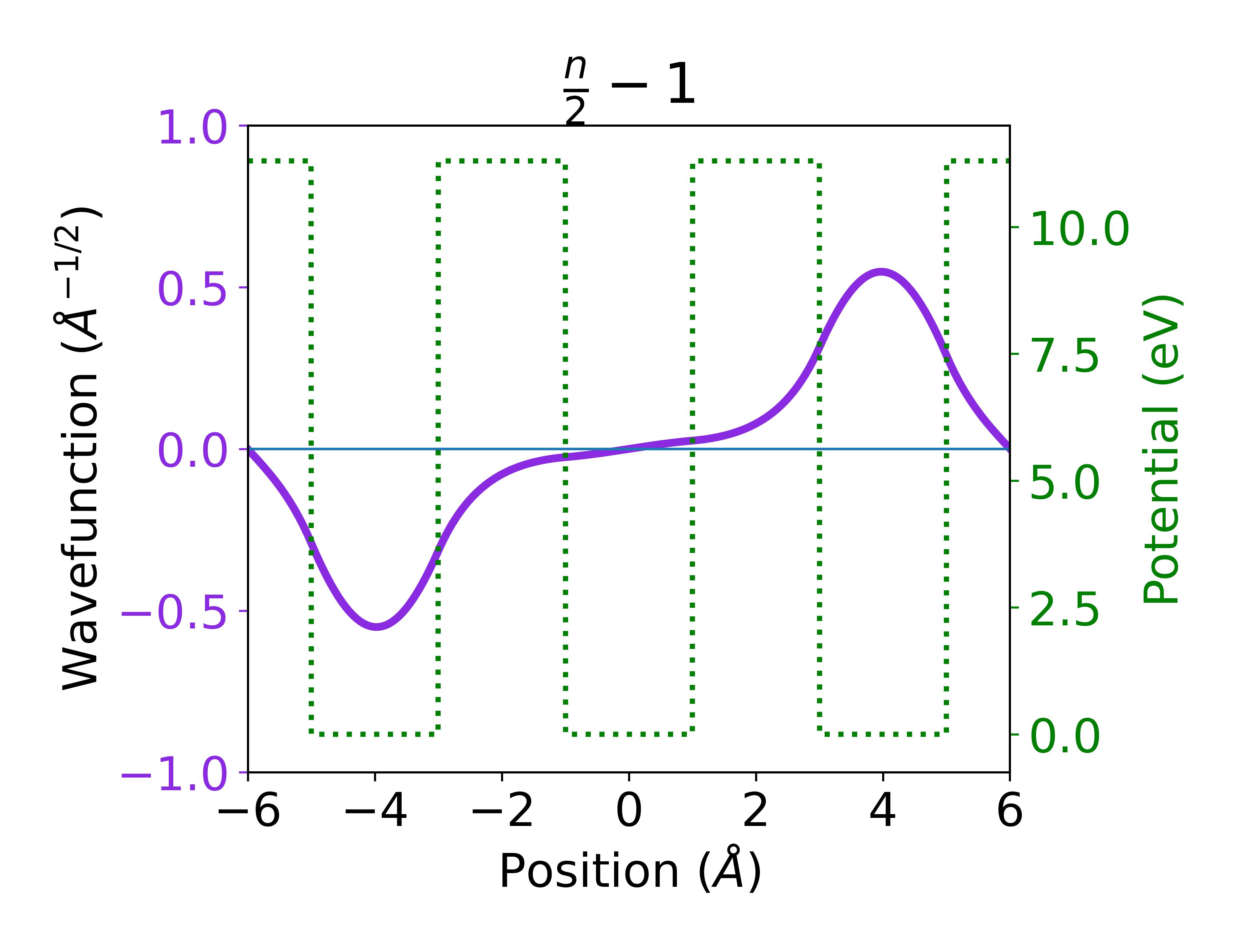}};
\node [anchor=north west] (imgB) at (0.40\linewidth,.58\linewidth){\includegraphics[width=0.49\linewidth]{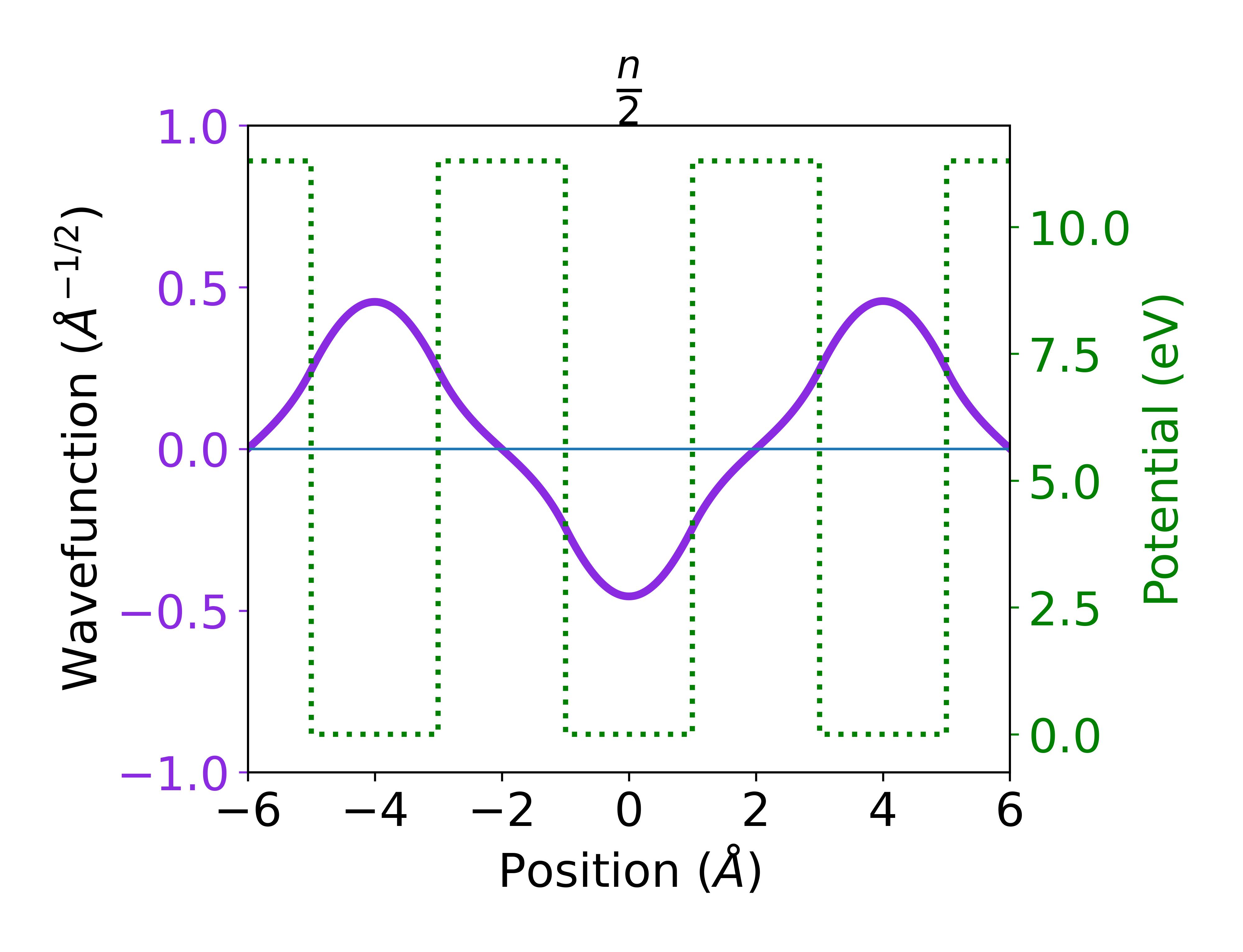}};    
\node [anchor=north west] (imgC) at (-0.10\linewidth,.2\linewidth){\includegraphics[width=0.48\linewidth]{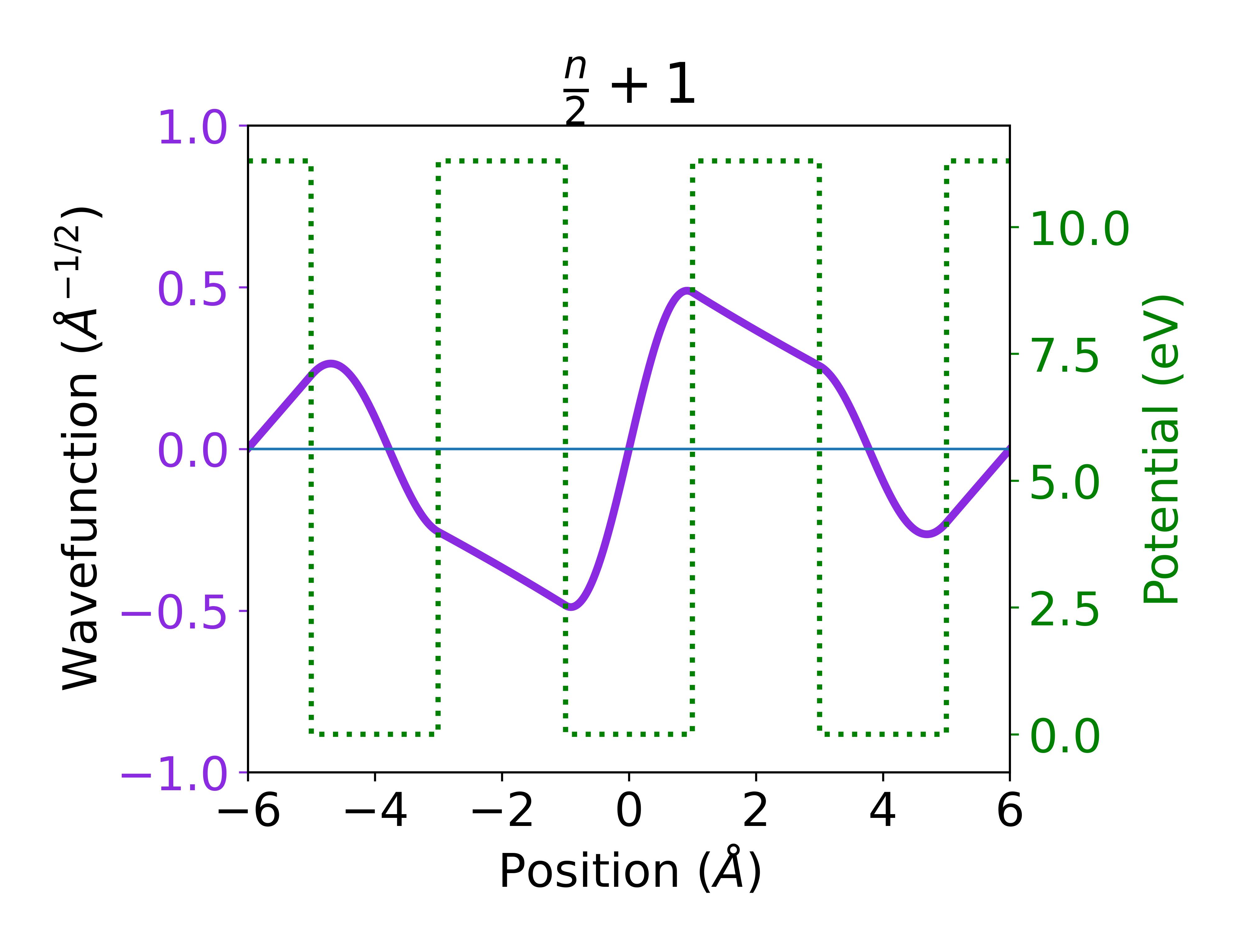}};
\node [anchor=north west] (imgD) at (0.40\linewidth,.2\linewidth){\includegraphics[width=0.48\linewidth]{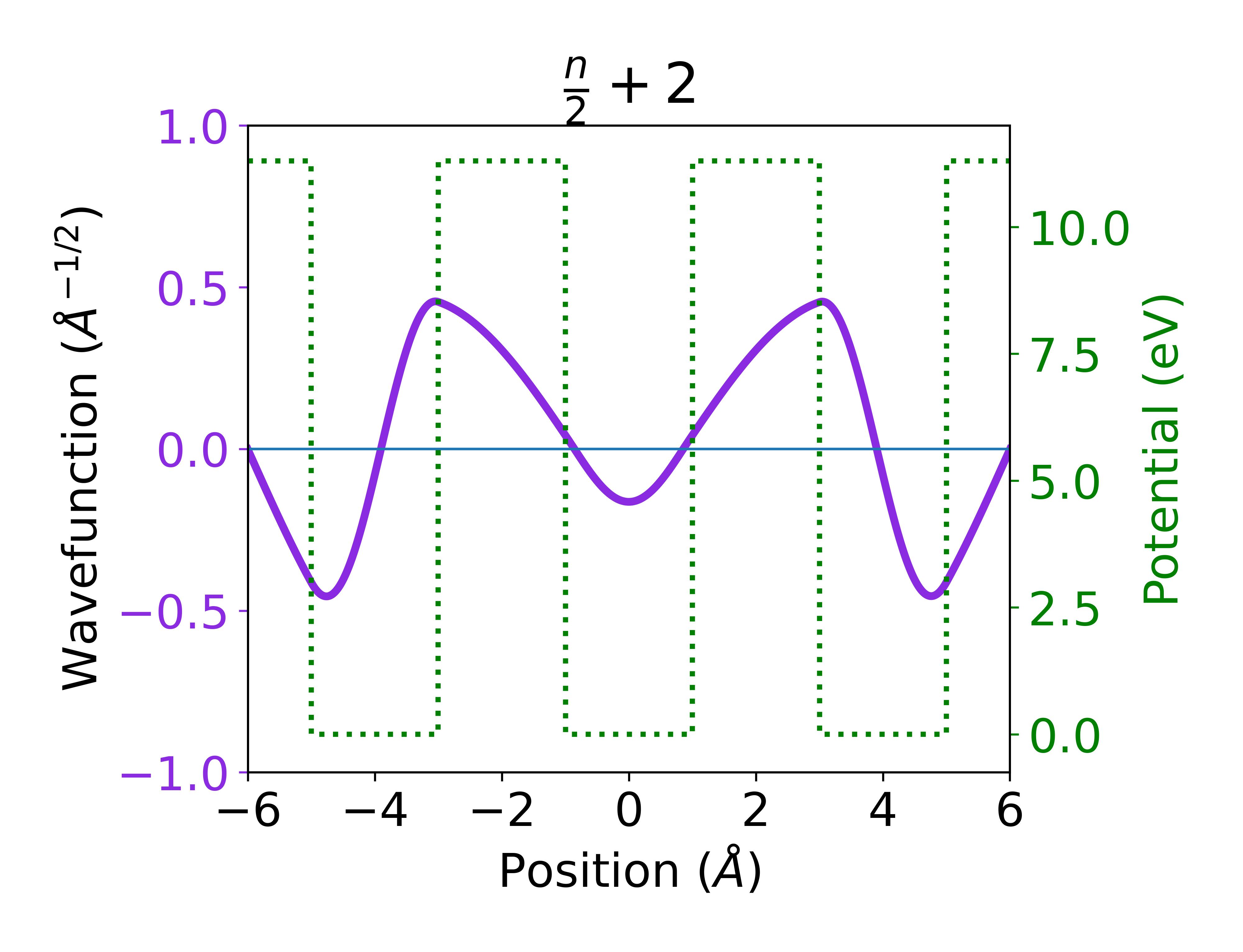}};
\draw [anchor=north west] (-0.10\linewidth, .58\linewidth) node {\textbf{(a)} };
\draw [anchor=north west] (0.40\linewidth, .58\linewidth) node {\textbf{(b)} };
\draw [anchor=north west] (-0.10\linewidth, .2\linewidth) node {\textbf{(c)} };
\draw [anchor=north west] (0.41\linewidth, .2\linewidth) node {\textbf{(d)} };
\end{tikzpicture}}
\caption{Wave functions for Well-Centered System (6e)}
\label{WF-Well}
 \end{figure*}

In this work, we focus on the well-centered system, for which one needs to subtract the $n/2$ state from the $n/2+1$ state, containing 2 and 3 nodes, respectively. These states can be seen in figure \ref{WF-Well}, which results in a band gap of 6.793 eV in the approach to the limit. However, this value is slightly below the periodic value.

\begin{figure}[h]
\centering
\includegraphics[width=.8\columnwidth]{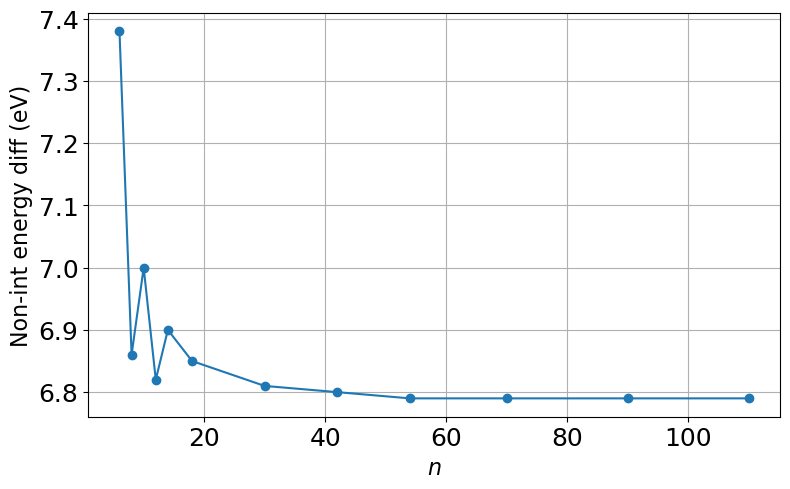} 
\caption{Bandgaps approaching periodicity for non-interacting well centered system}
\label{WBGVN}
\end{figure}

 We use the well-centered system because it uses the $n/2$ and the $n/2 +1$ states. 
We also did this analysis for the Barrier and Edge centered system with different results. The barrier centered system acted much like the well centered system where if the pattern was shifted from $4m +2$ to $4m$ the HOMO and LUMO changed states as seen below in  Fig. \ref{BarrierStates}. 
\begin{figure}[h]
\centering
\includegraphics[width=.8\columnwidth]{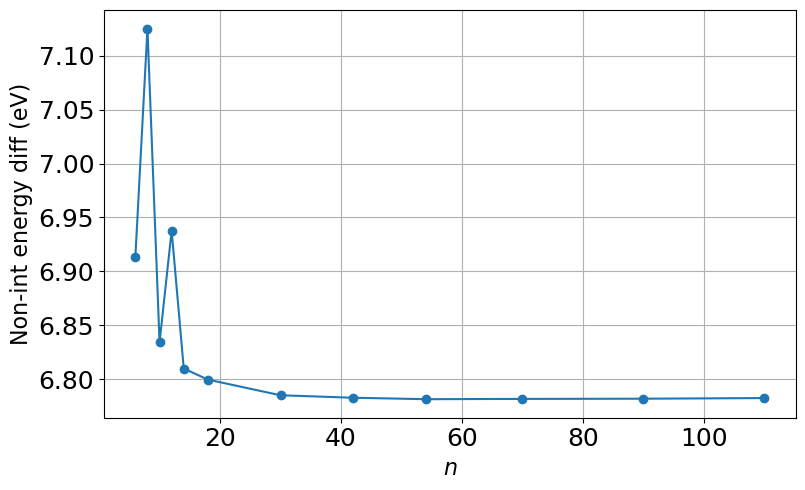} 
\caption{Bandgaps approaching periodicity for non-interacting barrier centered system}
\label{BBGVN}
\end{figure}

\begin{figure}[h]
\centering
\includegraphics[width=.8\columnwidth]{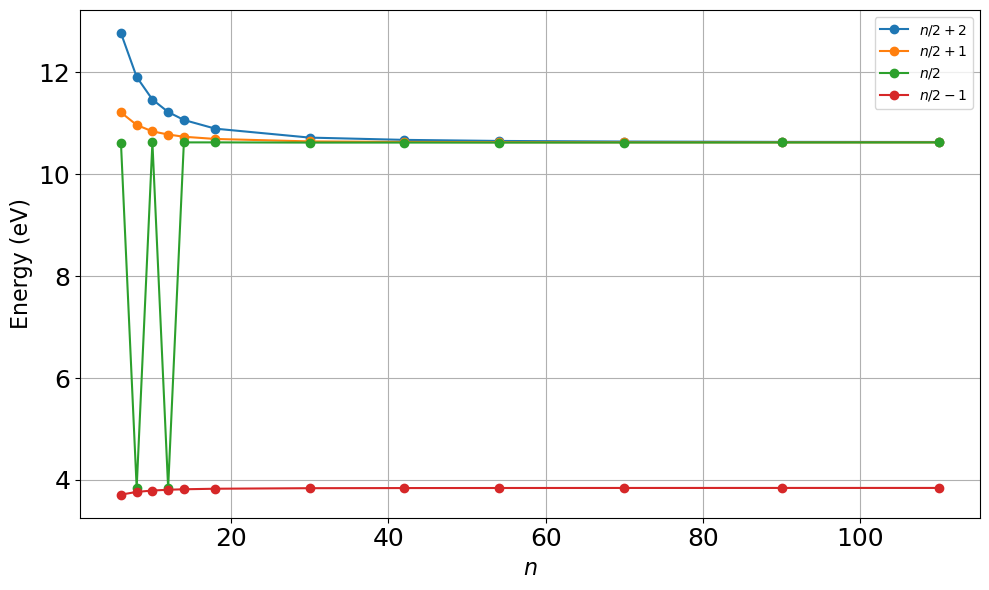} 
\caption{State energies of the approach to periodicity for non-interacting system in barrier centerd system. Degeneracy occurs when switching between the two electron count patterns. }
\label{BarrierStates}
\end{figure}

\begin{figure}[h]
\centering
\includegraphics[width=.8\columnwidth]{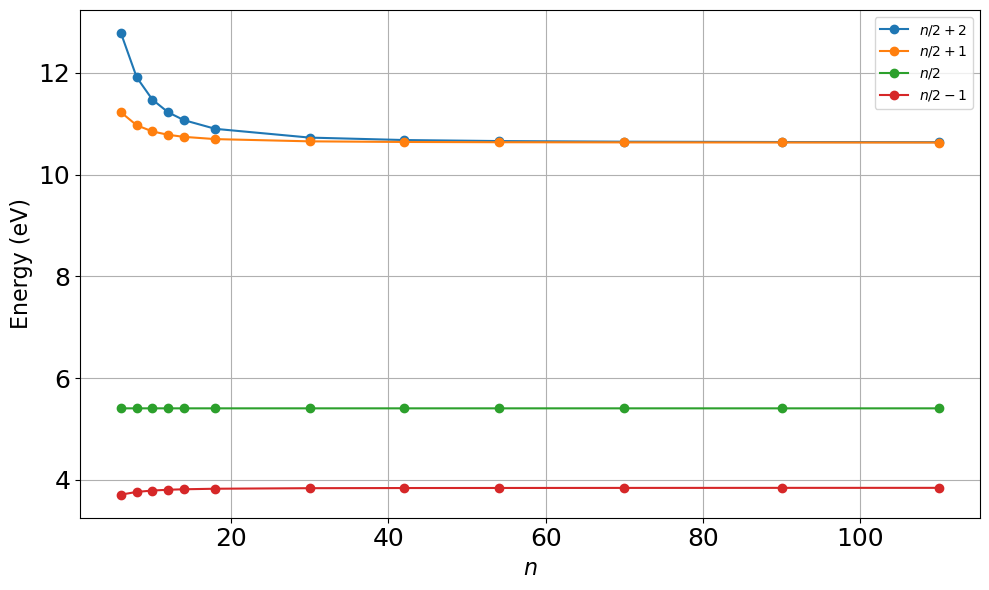} 
\caption{State energies of the approach to periodicity for non-interacting system in edge centered system, with $n/2$ state acting as edge state. }
\label{EdgeStates}
\end{figure}

However, in the case for the edge centered system this did not occur. Instead, despite running tests in both the $4m +2$ and $4m$ patterns, the HOMO and LUMO remained with the same state as can be seen in Fig. \ref{EdgeStates}.
\begin{figure}[h]
\centering
\includegraphics[width=.8\columnwidth]{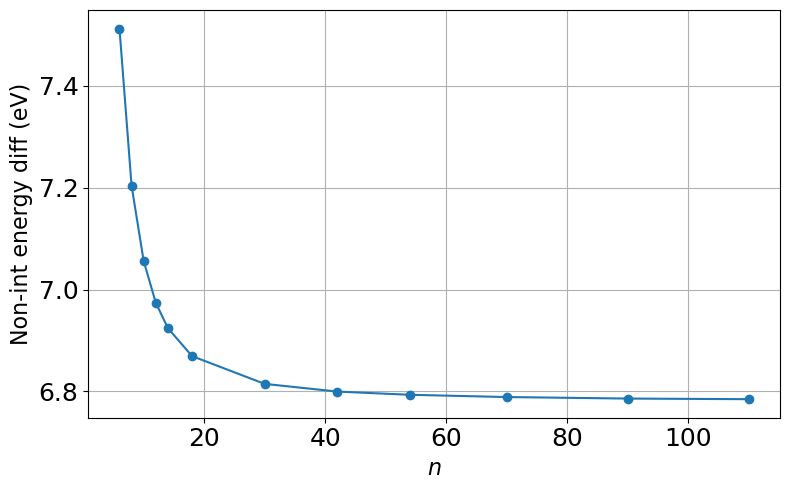} 
\caption{Bandgaps approaching periodicity for non-interacting edge centered system}
\label{EBGVN}
\end{figure}

\subsection{Interacting System}
For the data produced by the ensemble portion of the data, they seem to approach a specific value of 10.0 eV for the XC broken symmetry case\cite{HFC11}. This is the case for enforced symmetry values.  The first symmetry value that follows equation 42 in Ref, \cite{Remi} is as follows. It seems to be approaching a value of roughly 9.9 eV. 
\begin{equation}
\Omega_{1}^{\text{e}} = \epsilon_{n=N_e/2+1} - \epsilon_{n=N_e/2} -  E_{\text{Hxc}}[\rho_{\text{GS}}] + E_{\text{Hxc}}[\rho_{\alpha \alpha}].
\label{eqn:Omega1e}
\end{equation}

\begin{figure}[h]
\centering
\includegraphics[width=.8\columnwidth]{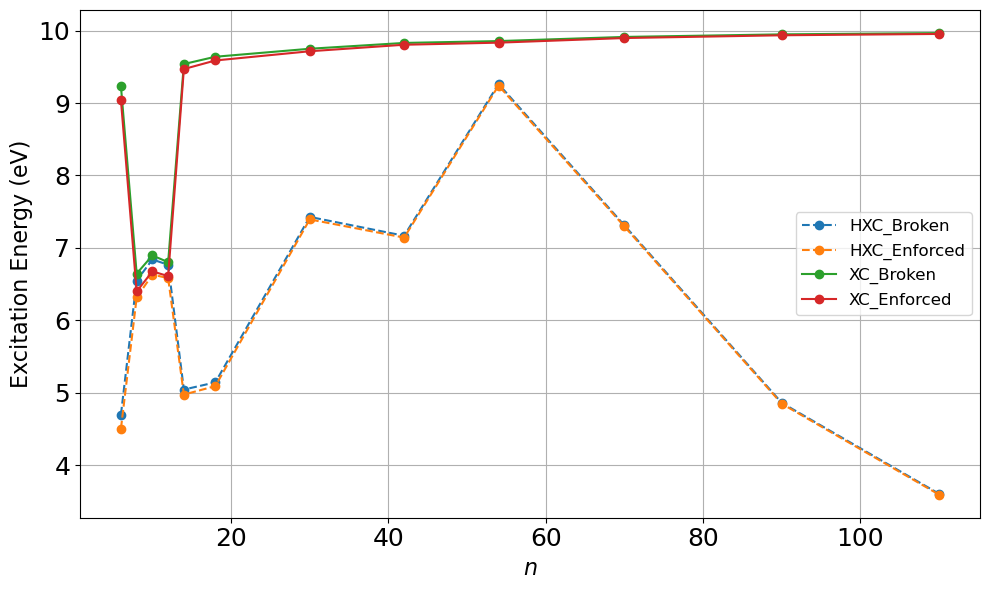} 
\caption{XC and HXC bandgaps as one approachs periodicity for both Broken and Enforced Symmetry well centered system.}
\label{XC+HXC}
\end{figure}

The second XC symmetry value which follows the broken symmetry system can be seen below. It also seems to be approaching a value of 10 eV. 
\begin{equation}
\Omega_{1}^{\text{b}} = \epsilon_{n=N_e/2+1} - \epsilon_{n=N_e/2} - E_{\text{Hxc}}[\rho_{\text{GS}}] + \frac{2}{3} E_{\text{Hxc}}[\rho_{\alpha \alpha}] + \frac{1}{3}E_{\text{Hxc}}[\rho_{\alpha\beta \pm \beta\alpha}]
\label{eqn:Omega1b}
\end{equation}

As for the HXC symmetry systems\cite{C96}, they do not seem to follow the same logic and approach a consistent value at periodicity. As seen in the figure \ref{XC+HXC}, they are not being as consistent as XC, which leads us to think that HXC not as good of an approximation method for this kind of calculation.  
We also note that at lower electron counts for both the XC and HXC system, they seem to shoot either up or down to inconsistent values. While phenomenon this is interesting to examine, it does not relate to our conclusions or results since it seems to be limited to those lower electron counts, so we will not be analyzing further. 
Finally, one can see that there is a jump in regard to the degeneracy of in the well centered and barrier centered systems. It can be seen that as those two systems transfer from the $4m +2$ to the $4m$ systems, that either the HOMO system becomes the LUMO system in the case of the Well centered, or vice versa in the case of the barrier centered. These degenerate states occur for an unknown reason. Lower states have been analyzed to see if the changing of the electron pattern adds or subtracts an additional state to explain this shift, but this was not found to be the case. 

\section{Conclusion}
Due to agreement between the independent particles (Kohn Sham) version of the system's and  periodic version’s bandgap in the limit, we know that as one approaches an infinitely large system, one can find out which finite states correspond to bulk band edges. We also confirm that the number of nodes remains consistent for finding the upper and lower bounds of the states for the bandgap regardless of whether the $4m$ or $4m+2$ centering is used, meaning that for bandgap calculations in different electron count centerings, finding the bandgap is node dependent not state dependent. 

As for the EDFT data, since both sets of data for XC approach a consistent bandgap, we find that it is promising to use EDFT to look at semiconductor systems. The EDFT calculations give a correction of the bandgap from the Kohn-Sham value of 6.8 eV to 10 eV, consistent with the expectation that the bandgap predicted by EDFT should be larger than the KS bandgap, and reasonable in magnitude for quasiparticle gap corrections to KS. The inconsistencies with the plots seem to be limited to the smaller electron counts, and therefore do not affect our final results.

\section{Acknowledgments}
G.K. was supported by the National Science Foundation (NSF) under grant (Award 2150531) for the Physics Research at UC Merced REU program. R.J.L. was supported by the NRT program Convergence of Nano-engineered Devices for Environmental and Sustainable Applications (CONDESA) under NSF award DGE-2125510. D.A.S. was supported by the U.S. Department of Energy, National Nuclear Security Administration, Minority Serving Institution Partnership Program, under Award DE-NA0003984. Computational resources were provided by the Pinnacles cluster at Cyber-Infrastructure and Research Technologies (CIRT), University of California, Merced, supported by the National Science Foundation Award OAC-2019144.

\section{Supplemental Materials}
\begin{table}[h]
    \centering
    \begin{tabular}{|c|c|}
        \hline
        \textbf{Term} & \textbf{Value} \\ \hline
        CalculationMode & gs \\ \hline
        ExtraStates & 10 \\ \hline
        PeriodicDimensions & 0 \\ \hline
        Dimensions & 1 \\ \hline
        TheoryLevel & independentparticles \\ \hline
        SpinComponents & spinpolarized \\ \hline
        FromScratch & yes \\ \hline
        n & electron count \\ \hline
        ConvEigenError & true \\ \hline
        preconditioner & no \\ \hline
        EigensolverMaxIter & 50 \\ \hline
        MaximumIter & -1 \\ \hline
        Radius & n*(a+b)/2 \\ \hline
        UnitsOutput & evangstrom \\ \hline
        Spacing & (a+b)/1000 \\ \hline
    \end{tabular}
    \caption{Definitions for Independent particles input file.}
    \label{tab:placeholder_label}
\end{table}

\begin{figure}[h]
\centering
\includegraphics[width=.8\columnwidth]{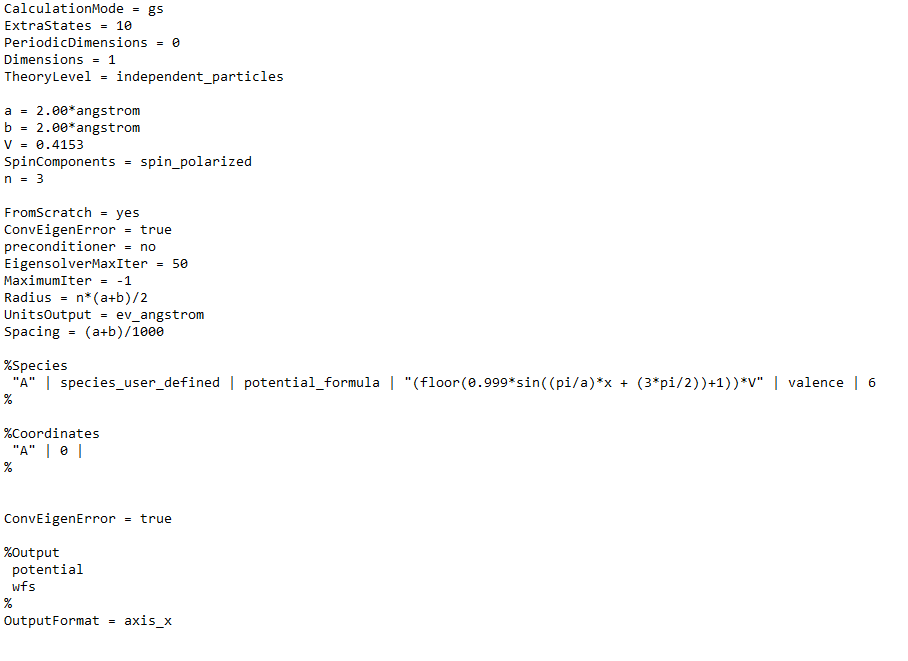} 
\caption{Example of well centered 6e input file for reference of variables used in table 1}
\label{Inp}
\end{figure}

\begin{table}[h]
    \centering
    \begin{tabular}{|c|c|c|c|c|c|}
        \hline
        \textbf{State} & \textbf{$4m + 2$ Well (6e) } & \textbf{$4m$ Well (8e) } & \textbf{$4m + 2$ Edge (6e) } & \textbf{$4m + 2$ Barrier (6e) } & \textbf{Periodic}\\ \hline
        $n/2 + 2$ & LUMO + 1 & LUMO + 2 & LUMO + 1 & LUMO + 2 & LUMO + 2 \\ \hline 
        $n/2 + 1$ & LUMO & LUMO + 1 & LUMO & LUMO + 1 & LUMO + 1 \\ \hline
        $n/2$     & HOMO & LUMO & EDGE & LUMO & LUMO\\ \hline
        $n/2 - 1$ & HOMO - 1 & HOMO & HOMO & HOMO & HOMO \\ \hline
         
    \end{tabular}
    \caption{Definitions of states and what they mean in electron counts for Well-centered system}
    \label{StateClasstable}
\end{table}

\begin{table}[htbp]
\centering
\caption{Combined definitions and wave functions for a different-centered system. With each HOMO state having 0 nodes per well, and each LUMO state having 1 node per well. The periodic system at k point 12.}
\label{tab:aligned_states}

\resizebox{\textwidth}{!}{%
\begin{tabular}{>{\centering\arraybackslash}p{2.4cm} | *{5}{>{\centering\arraybackslash}p{3.2cm}} }
\toprule
 & \textbf{$4m + 2$ Well} & \textbf{$4m$ Well} & \textbf{$4m + 2$ Edge} & \textbf{$4m + 2$ Barrier} & \textbf{Periodic  k = 0.5} \\
\midrule

\textbf{HOMO – 1} &
\makecell{\adjustbox{valign=m}{\includegraphics[width=\linewidth,height=3cm,keepaspectratio]{_-1_W.jpg}}} &
 & & & \\[0.5em]
\midrule

\textbf{HOMO} &
\makecell{\adjustbox{valign=m}{\includegraphics[width=\linewidth,height=3cm,keepaspectratio]{_0_W.jpg}}} &
\makecell{\adjustbox{valign=m}{\includegraphics[width=\linewidth,height=3cm,keepaspectratio]{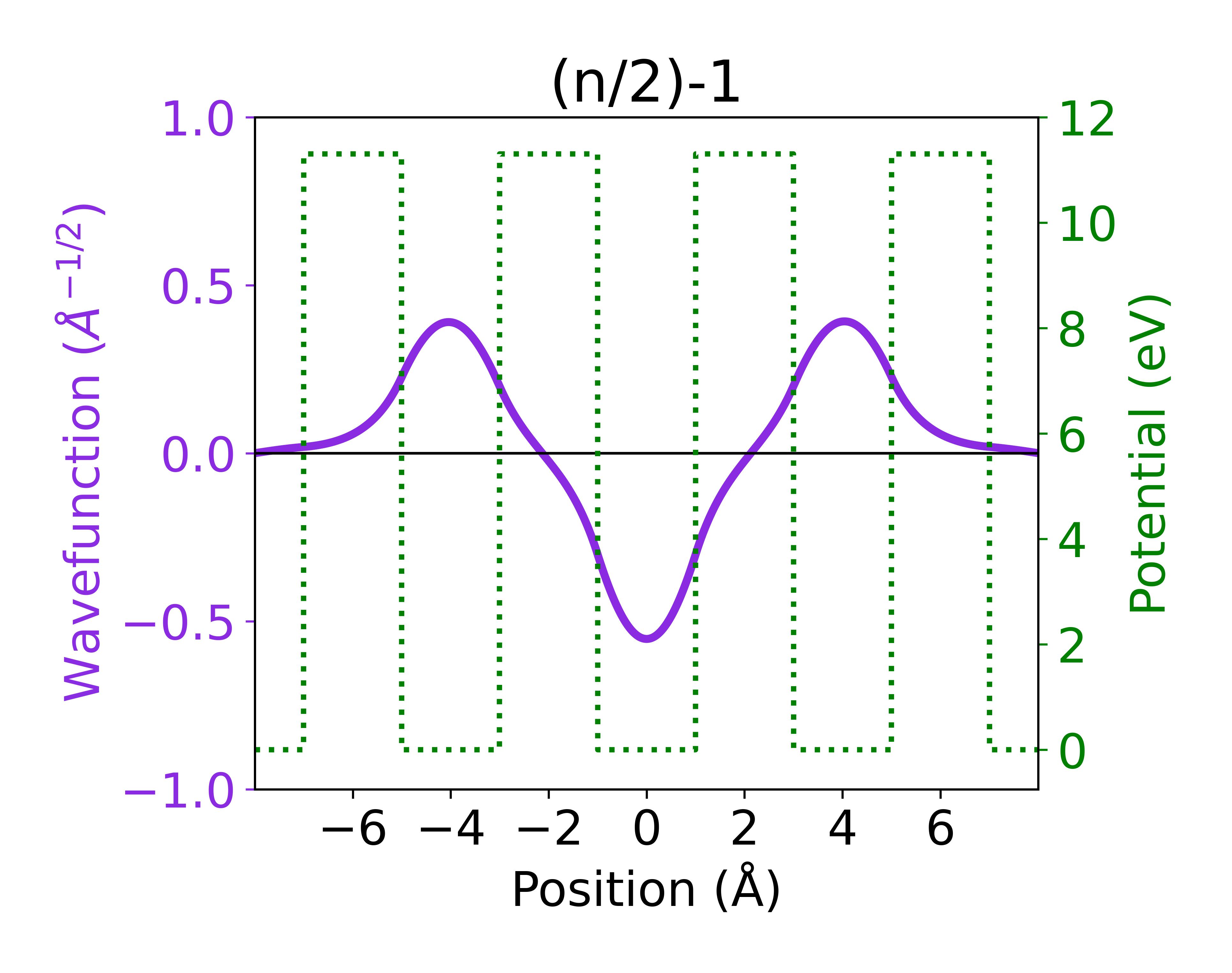}}} &
\makecell{\adjustbox{valign=m}{\includegraphics[width=\linewidth,height=3cm,keepaspectratio]{_-1_E.jpg}}} &
\makecell{\adjustbox{valign=m}{\includegraphics[width=\linewidth,height=3cm,keepaspectratio]{_-1_B.jpg}}} &
\makecell{\adjustbox{valign=m}{\includegraphics[width=\linewidth,height=3cm,keepaspectratio]{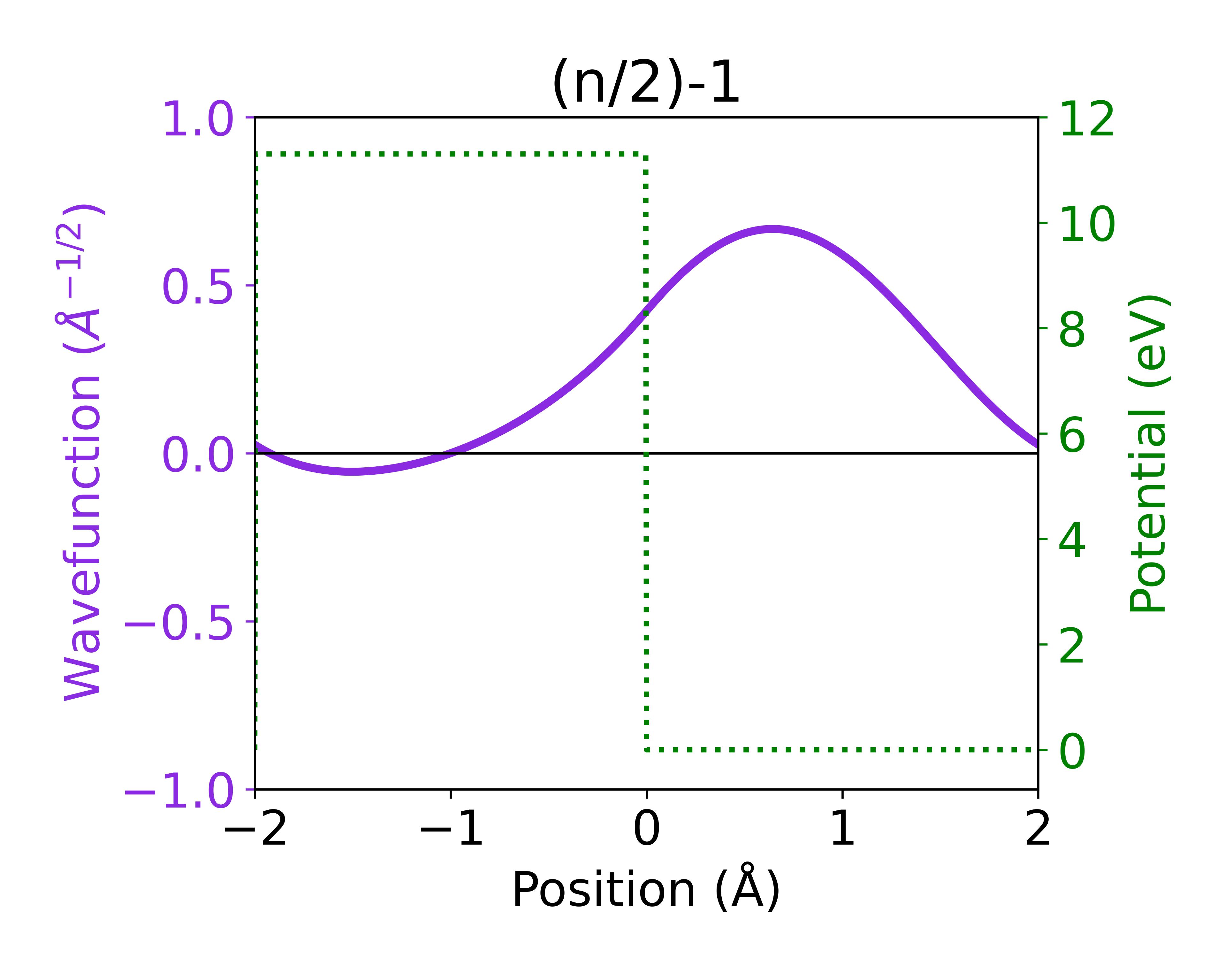}}} \\[0.5em]
\midrule

\textbf{EDGE} &
 & & 
\makecell{\adjustbox{valign=m}{\includegraphics[width=\linewidth,height=3cm,keepaspectratio]{_0_E.jpg}}} &
 & \\[0.5em]
\midrule

\textbf{LUMO} &
\makecell{\adjustbox{valign=m}{\includegraphics[width=\linewidth,height=3cm,keepaspectratio]{_+1_W.jpg}}} &
\makecell{\adjustbox{valign=m}{\includegraphics[width=\linewidth,height=3cm,keepaspectratio]{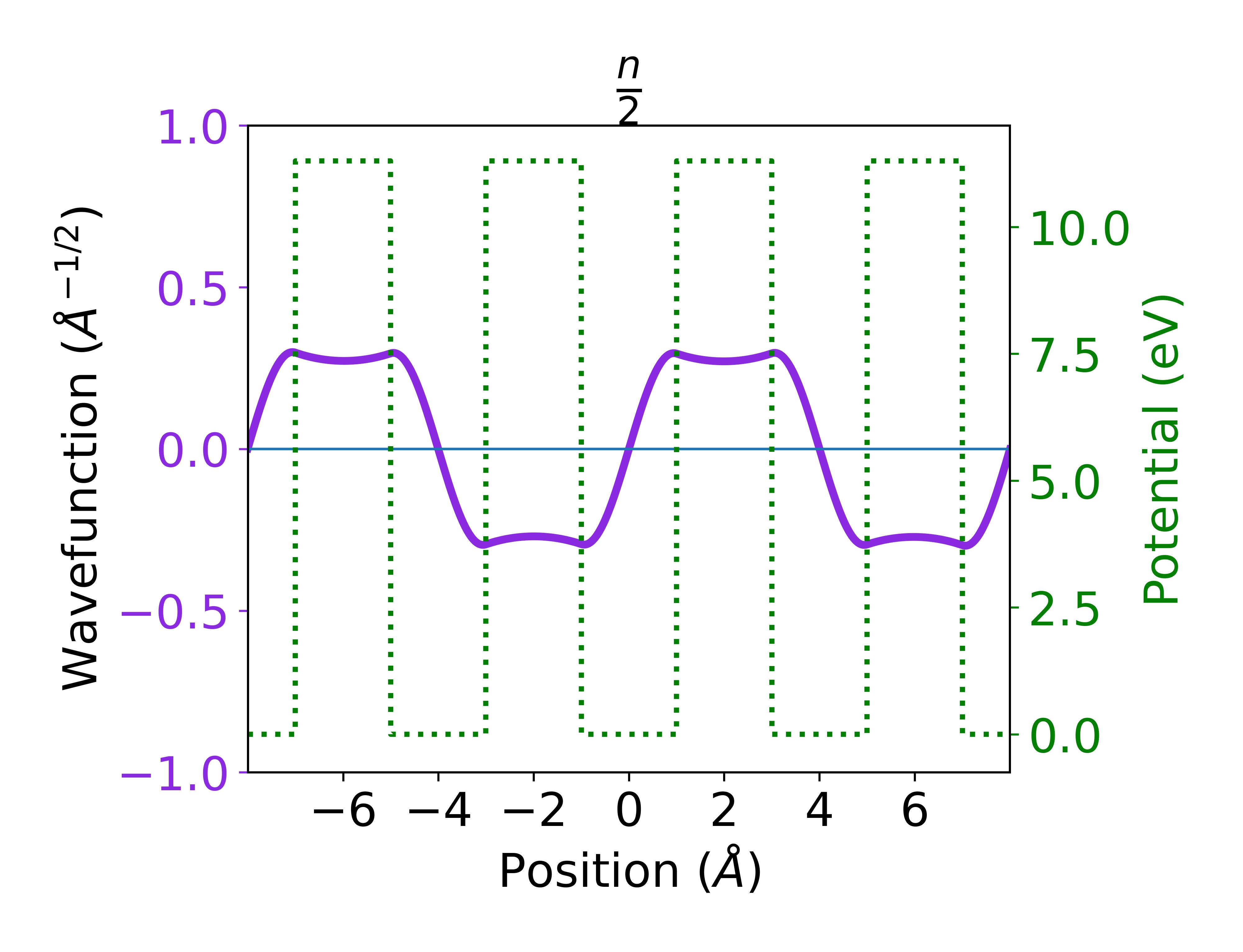}}} &
\makecell{\adjustbox{valign=m}{\includegraphics[width=\linewidth,height=3cm,keepaspectratio]{_+1_E.jpg}}} &
\makecell{\adjustbox{valign=m}{\includegraphics[width=\linewidth,height=3cm,keepaspectratio]{_0_B.jpg}}} &
\makecell{\adjustbox{valign=m}{\includegraphics[width=\linewidth,height=3cm,keepaspectratio]{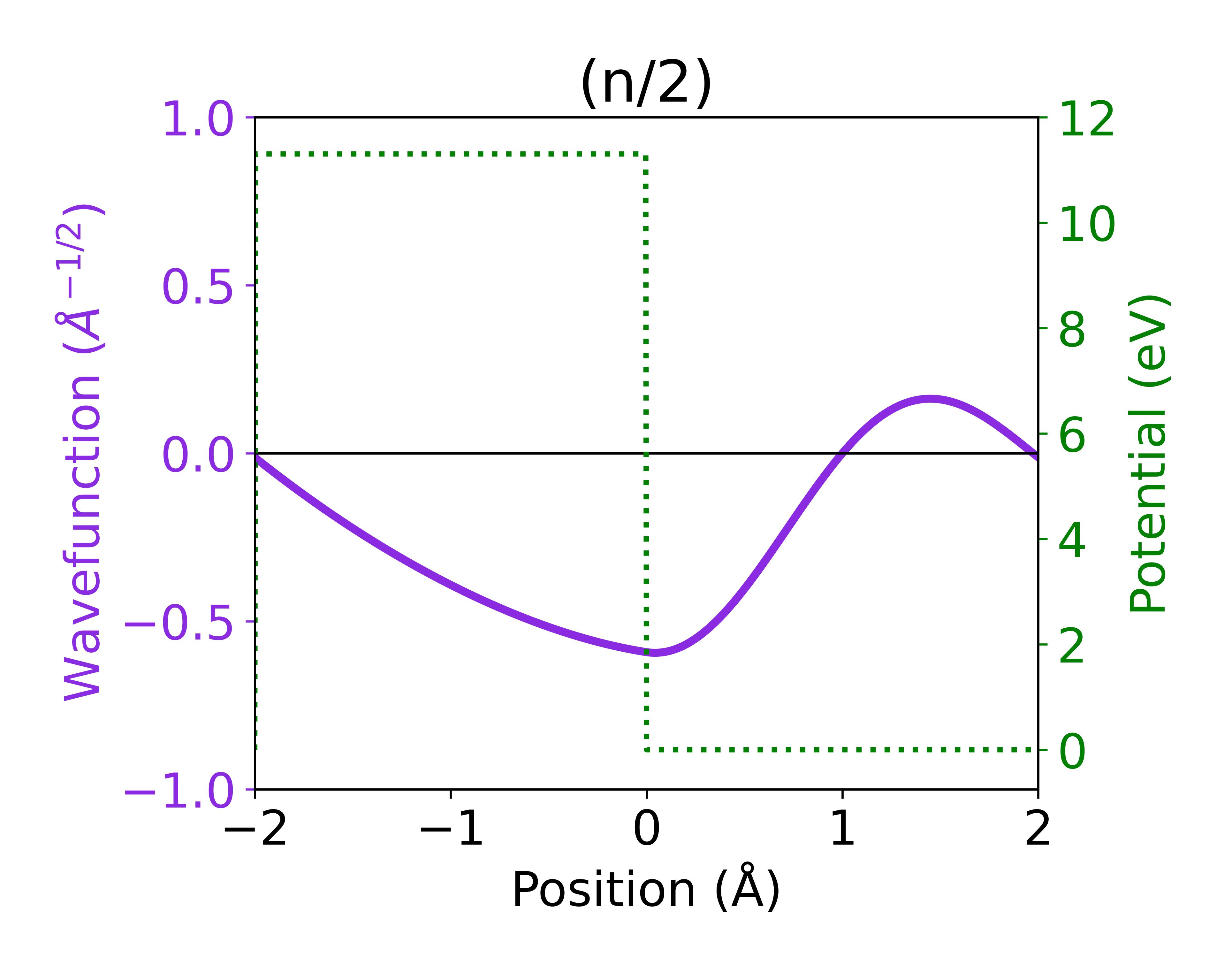}}} \\[0.5em]
\midrule

\textbf{LUMO + 1} &
\makecell{\adjustbox{valign=m}{\includegraphics[width=\linewidth,height=3cm,keepaspectratio]{_+2_W.jpg}}} &
\makecell{\adjustbox{valign=m}{\includegraphics[width=\linewidth,height=3cm,keepaspectratio]{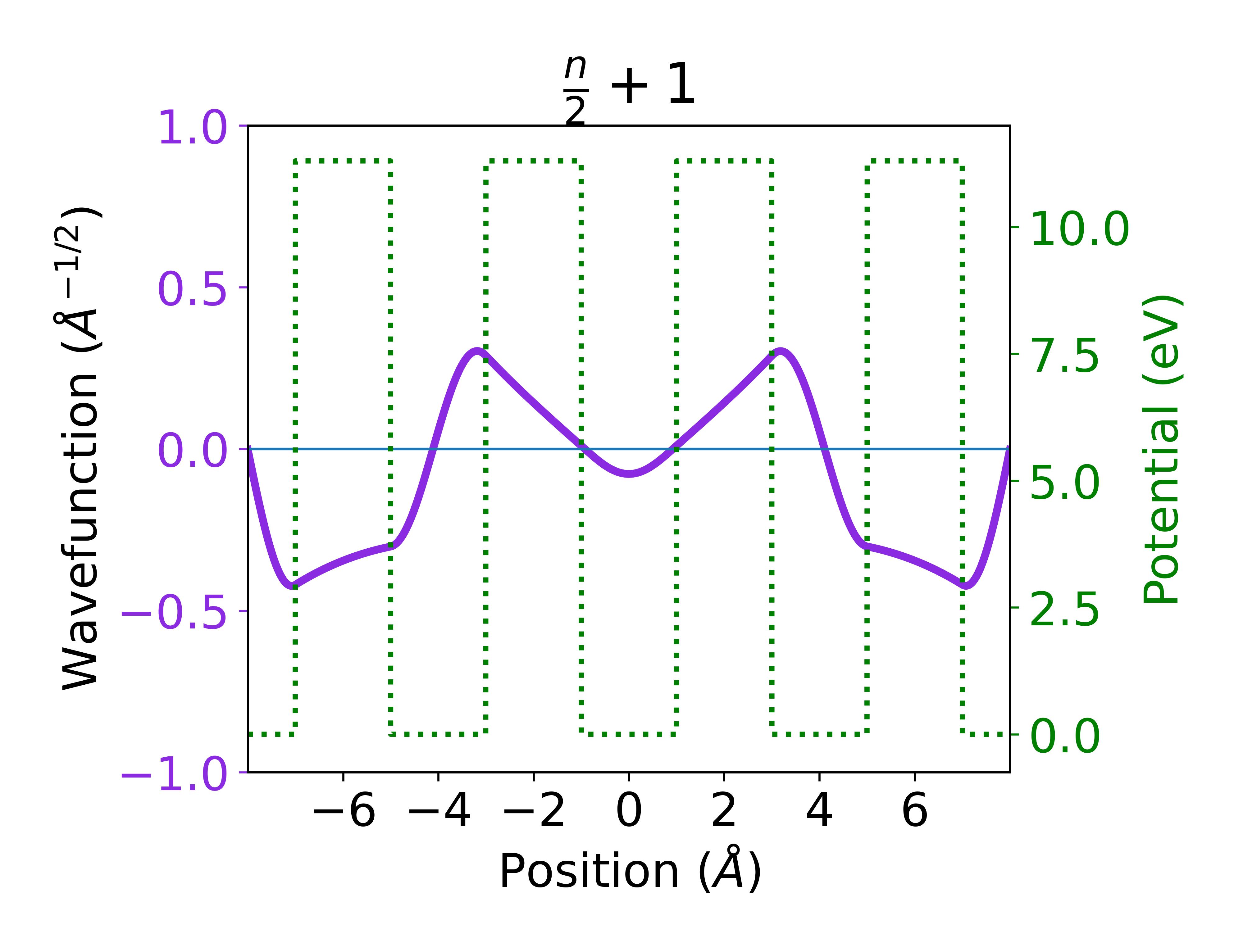}}} &
\makecell{\adjustbox{valign=m}{\includegraphics[width=\linewidth,height=3cm,keepaspectratio]{_+2_E.jpg}}} &
\makecell{\adjustbox{valign=m}{\includegraphics[width=\linewidth,height=3cm,keepaspectratio]{_+1_B.jpg}}} &
\makecell{\adjustbox{valign=m}{\includegraphics[width=\linewidth,height=3cm,keepaspectratio]{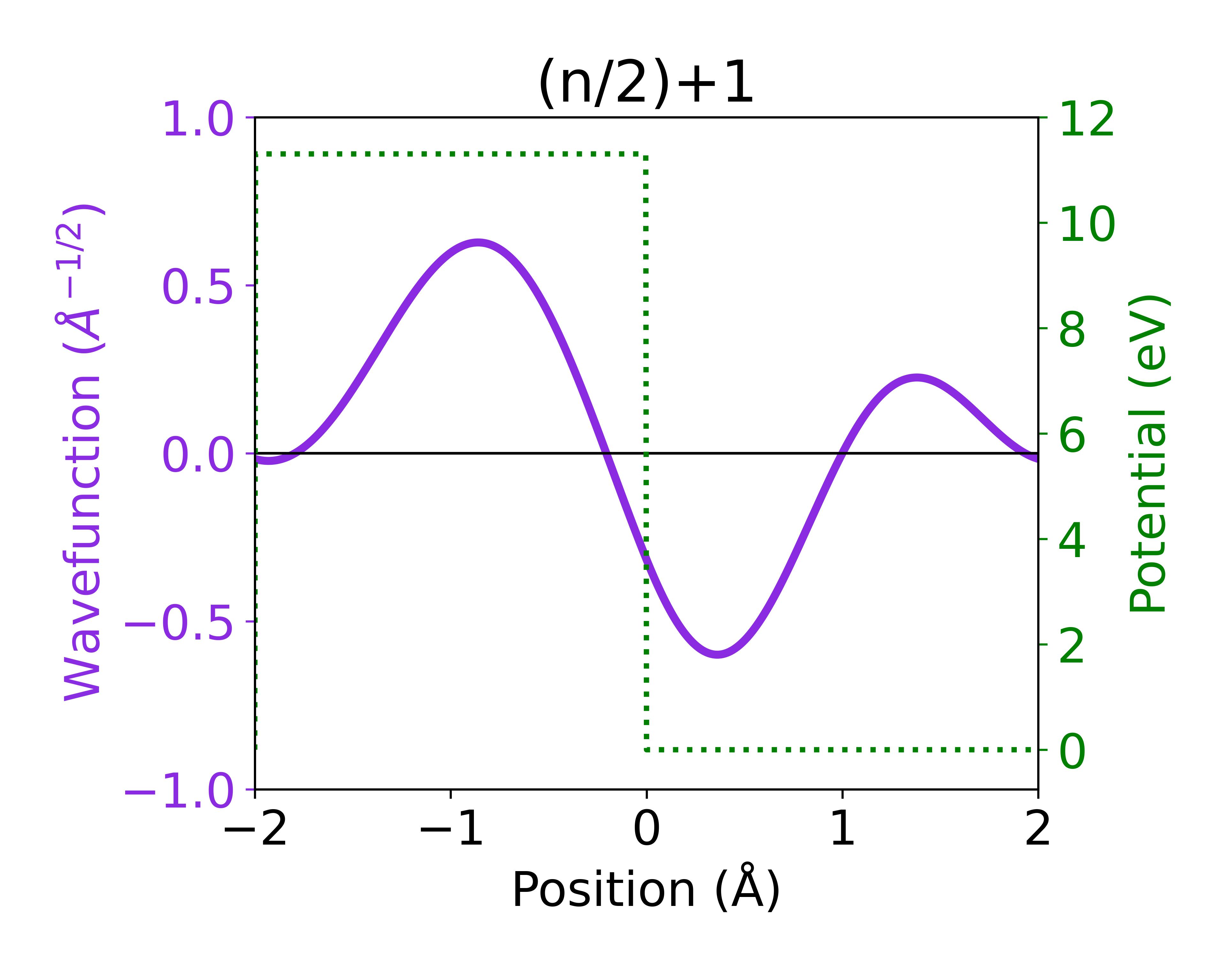}}} \\[0.5em]
\midrule

\textbf{LUMO + 2} &
 & 
\makecell{\adjustbox{valign=m}{\includegraphics[width=\linewidth,height=3cm,keepaspectratio]{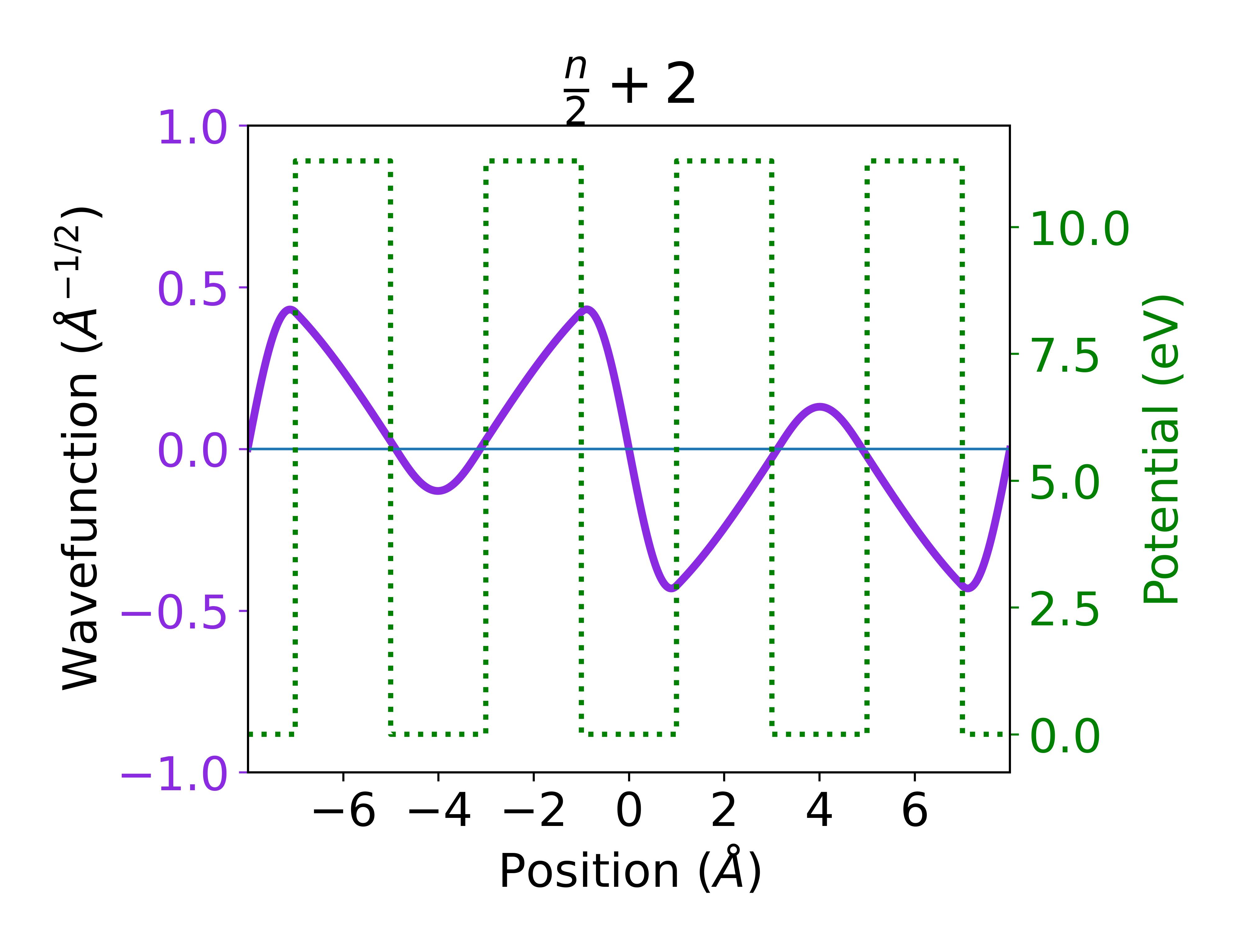}}} &
 & 
\makecell{\adjustbox{valign=m}{\includegraphics[width=\linewidth,height=3cm,keepaspectratio]{_+2_B.jpg}}} &
\makecell{\adjustbox{valign=m}{\includegraphics[width=\linewidth,height=3cm,keepaspectratio]{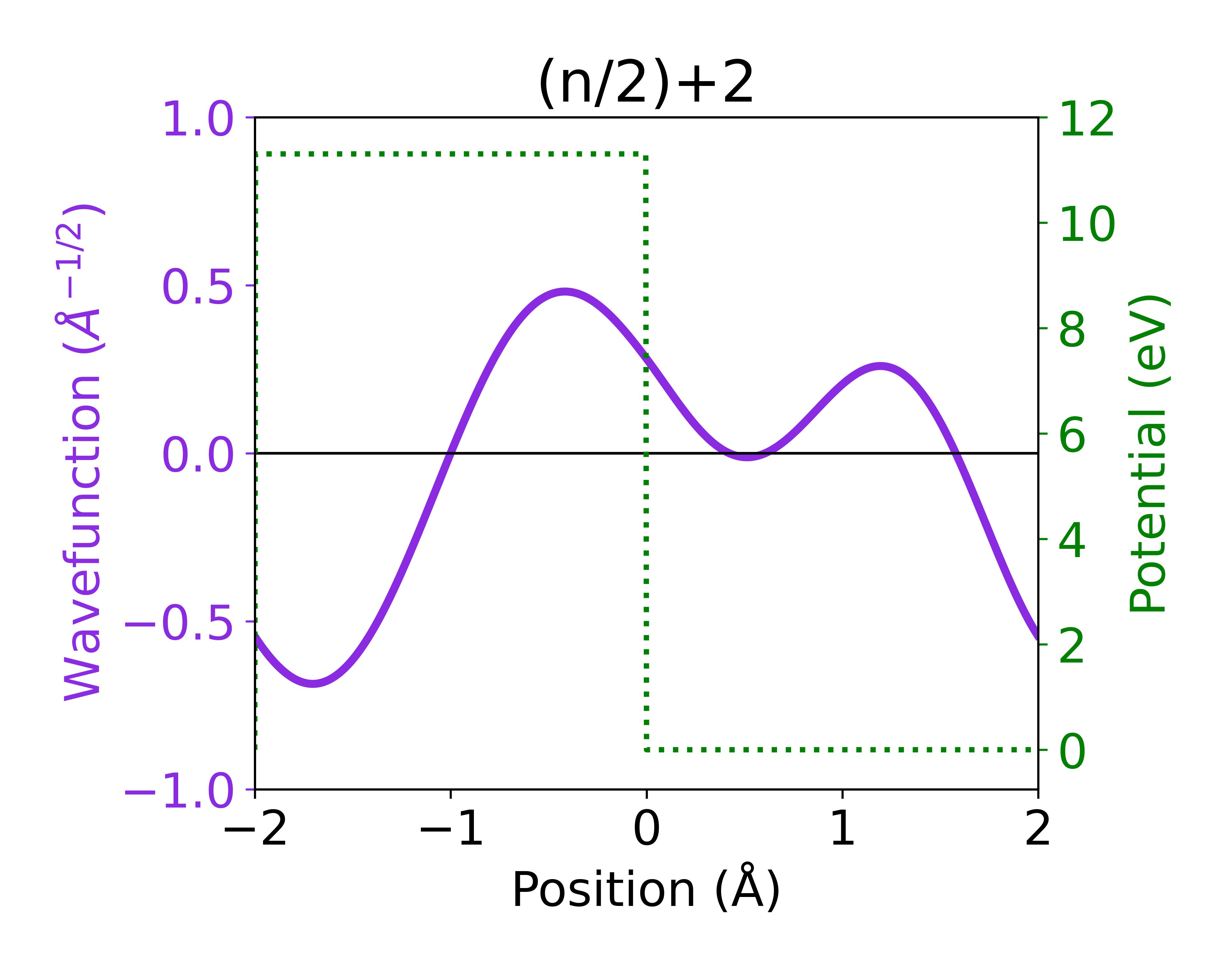}}} \\[0.5em]
\bottomrule
\end{tabular}
} %
\end{table}
\begin{table}[h]
    \centering
    \begin{tabular}{|c|c|}
        \hline
        \textbf{Label} & \textbf{Bandgap (eV) } \\ \hline
        Periodic & 6.793  \\ \hline
        Well-centered Independent Particles & 6.785  \\ \hline
        Well-centered Enforced XC & 9.954  \\ \hline
        Well-centered Enforced HXC & 3.591\\ \hline  
        Well-centered Broken XC &  9.966 \\ \hline
        Well-centered Broken HXC & 3.603 \\ \hline
        Barrier-centered Independent Particles & 6.782  \\ \hline
        Edge-centered Independent Particles & 6.784  \\ \hline
    \end{tabular}
    \caption{Bandgaps for different centered systems including periodic, independent particles, and EDFT calculations with confidence up to the thousandths place. Note Values for the HXC bandgaps are not a converged value due to lack of convergence, but simply the value outputted at 110 e}
    \label{tab:bandgaps}
\end{table}
\bibliography{aipsamq1p}

\end{document}